\documentclass[sigconf]{acmart}
\copyrightyear{2020}
\acmYear{2020}
\setcopyright{acmcopyright}\acmConference[KDD '20]{Proceedings of the 26th ACM SIGKDD Conference on Knowledge Discovery and Data Mining}{August 23--27, 2020}{Virtual Event, CA, USA}
\acmBooktitle{Proceedings of the 26th ACM SIGKDD Conference on Knowledge Discovery and Data Mining (KDD '20), August 23--27, 2020, Virtual Event, CA, USA}
\acmPrice{15.00}
\acmDOI{10.1145/3394486.3403198}
\acmISBN{978-1-4503-7998-4/20/08}

\usepackage{bm,nicefrac}
\usepackage{microtype}
\usepackage{graphicx}
\usepackage{subfigure}
\usepackage{amsmath}
\usepackage{color}
\usepackage{url}

\newcommand{\tf}{\texttt{TF-Net}}
\newcommand{\ra}[1]{\renewcommand{\arraystretch}{#1}}

\begin{document}
\fancyhead{}
\title{Towards physics-informed deep learning \\ for turbulent flow prediction}

\author{Rui Wang}
\authornote{part of the work done while interning at Lawrence Berkeley National Laboratory}
\affiliation{%
  \institution{Northeastern University}
  \city{Boston}
  \country{MA}}
  \email{wang.rui4@northeastern.edu}

\author{Karthik Kashinath}
\affiliation{%
  \institution{Lawrence Berkeley Lab}
  \city{Berkeley}
  \country{CA}}
  \email{kkashinath@lbl.gov}

\author{Mustafa Mustafa}
\affiliation{%
  \institution{Lawrence Berkeley Lab}
  \city{Berkeley}
  \country{CA}}
  \email{mmustafa@lbl.gov}

\author{Adrian Albert}
\affiliation{%
   \institution{Lawrence Berkeley Lab}
  \city{Berkeley}
  \country{CA}}
  \email{aalbert@lbl.gov}
  
\author{Rose Yu}
\affiliation{%
  \institution{Northeastern University}
  \city{Boston}
  \country{MA}}
  \email{roseyu@northeastern.edu}

\renewcommand{\shortauthors}{Wang, et al.}

\begin{abstract}
While deep learning has shown tremendous success in a wide range of domains, it remains a grand challenge to incorporate physical principles in a systematic manner to the design, training and inference of such models. In this paper, we aim to predict turbulent flow by learning its highly nonlinear dynamics from spatiotemporal velocity fields of large-scale fluid flow simulations of relevance to turbulence modeling and climate modeling. We adopt a hybrid approach by  marrying two well-established turbulent flow simulation techniques with deep learning. Specifically, we  introduce trainable spectral filters in a coupled model of Reynolds-averaged Navier-Stokes (RANS) and Large Eddy Simulation (LES), followed by a specialized U-net for prediction.  Our approach, which we call Turbulent-Flow Net (\tf{}), is grounded in a principled physics model, yet offers the flexibility of learned representations. We compare our model, \tf{}, with state-of-the-art baselines and observe significant reductions in error for predictions  $60$  frames ahead. Most importantly, our method predicts physical fields that obey desirable physical characteristics, such as conservation of mass, whilst faithfully emulating the turbulent kinetic energy field and spectrum, which are critical for accurate prediction of turbulent flows. 
\end{abstract}





\maketitle

\section{Introduction}
Modeling the dynamics of large-scale spatiotemporal data  over a wide range of spatial and temporal scales is a fundamental task in science and engineering (e.g., hydrology,
solid mechanics, chemistry kinetics). Computational fluid dynamics (CFD) is at the heart of climate modeling and has direct implications for understanding and predicting climate change.  However, the current paradigm in atmospheric CFD is purely \textit{physics-based}: known physical laws encoded in systems of coupled partial differential equations (PDEs) are solved over space and time via numerical differentiation and integration schemes. These methods are tremendously computationally-intensive, requiring significant computational resources and expertise. Recently,  \textit{data-driven} methods, including deep learning, have demonstrated great success in the automation, acceleration, and streamlining of highly compute-intensive workflows for science \cite{prabhat_nature_2019}. But existing deep learning methods are mainly statistical with little or no underlying physical knowledge incorporated, and are yet to be proven to be successful in capturing and predicting accurately the properties of complex physical systems. 


Developing deep learning methods that can incorporate physical laws in a systematic manner is a key element in advancing AI for physical sciences \cite{mlfm}.  Recently, several studies in data science and computational mathematics communities have attempted incorporating physical knowledge  into deep learning, an area coined as ``physics-informed deep learning''. For example,  \cite{PDE-CDNN} proposed a warping scheme to predict the sea surface temperature, but only considered the linear advection-diffusion equation. \cite{Anuj2017PGNN,jia2019physics} proposed physics guided neural networks to model temperature of lakes by explicitly regularising the loss function with physical constraints. Still, regularization is quite ad-hoc and it is  difficult to tune  the hyper-parameters of the regularizer. \cite{xie2018tempogan} and \cite{eulerian} developed deep learning models in the context of fluid flow animation, where physical consistency is less critical. \cite{jinlong_2019, constrain1} introduced statistical and physical constraints in the loss function to regularize the predictions for emulating physical simulations. However, their studies only focused on \textit{spatial} modeling without \textit{temporal} dynamics.  

In this paper, we investigate the  problem of predicting the evolution of spatiotemporal turbulent flow, governed by the high-dimensional non-linear Navier-Stokes equations. In contrast to modeling sea surface temperature or lake dynamics, turbulent flow is highly chaotic. The temporal evolution of the turbulent flow is exceedingly sensitive to initial condition and there is no analytical theory to characterize its dynamics. 
Furthermore, turbulent flow demonstrate multiscale behavior where chaotic motion of the flow is forced at different length and time scale. Additionally, high-resolution turbulent flow leads to   high-dimensional forecasting problem. For example,  a discretized $128 \times 128 \times 100$ velocity field has $10^6$ dimensions, which requires a large number of training data, and is prone to error propagation.

We propose a hybrid learning paradigm that unifies turbulence modeling and deep  learning (DL). We develop a novel deep learning model, Turbulent-Flow Net (\tf{}), that enhances the capability of predicting complex turbulent flows with deep neural networks.  \tf{} applies scale separation to model different ranges of scales of the  turbulent flow individually.  Building upon a promising and popular CFD technique, the  RANS-LES coupling approach \cite{turb3}, our model replaces \textit{a priori} spectral filters with trainable convolutional layers. We decompose the turbulent flow into three components, each of which is approximated by a specialized U-net to preserve invariance properties. To the best of our knowledge, this is the first hybrid framework of its kind for predicting turbulent flow. We compare our method with state-of-the-art baselines for forecasting velocity fields up to 60 steps ahead given the history. We observe that \tf{} is capable of generating both accurate and physically meaningful predictions that preserve critical quantities of relevance. 

In summary, our contributions are as follows:
\begin{enumerate}
\item We study the challenging task of turbulent flow prediction as a test bed to investigate incorporating physics knowledge into deep learning in a principled fashion. 
\item We propose a novel hybrid learning framework, \tf{}, that unifies a popular CFD technique,  RANS-LES coupling, with custom-designed deep neural networks. 
\item When evaluated on turbulence simulations, \tf{} achieves 11.1\% reduction in prediction RMSE, 30.1\% improvement in the energy spectrum, 21\%  turbulence kinetic energy RMSEs and 64.2\% reduction of flow divergence in difference from the target, compared to the best baseline.
\end{enumerate}

\section{Related work}
\paragraph{\normalfont\textbf{Spatiotemporal Forecasting}} Modeling the  spatiotemporal dynamics of a system in order to forecast the future is of critical importance in fields as diverse as physics, economics, and neuroscience \cite{strogatz2018nonlinear, rui2020symmetry}. Methods in dynamical system literature from physics \cite{izhikevich2007dynamical} to  neuroscience \cite{wainwright2005dynamical} describe the spatiotemporal dynamics with differential equations and are  \textit{physics-based}. They often cannot be solved analytically and are difficult to simulate numerically due to high sensitivity to initial conditions.  In data mining, most work  on spatiotemporal forecasting has been purely \textit{data-driven} where complex deep learning models are learned directly from data without explicitly enforcing physical constraints, e.g. \cite{xingjian2015convolutional, li2018diffusion, yi2018deep,yao2018deep}. A few recent works   \cite{Anuj2017PGNN,jia2019physics} tried to incorporate physical knowledge into deep learning by explicitly regularising the loss function with physical constraints. Still, regularization is quite ad-hoc and it is  difficult to tune  the hyper-parameters of the regularizer.
Perhaps most related to ours is a hybrid framework in \cite{HybridNet} which aims to predict the evolution of the external forces/perturbations but they did not try modeling turbulence.


\paragraph{\normalfont\textbf{Turbulence Modeling}} Recently, machine learning models, especially DL models have been used to accelerate and improve the simulation of turbulent flows. For example,  \cite{ling2016reynolds, fang2018deep} studied tensor invariant neural networks to learn the Reynolds stress tensor while preserving Galilean invariance, but Galilean invariance only applies to flows without external forces. In our case, RBC flow has gravity as an external force. Most recently, \cite{kim2019deep} studied unsupervised generative modeling of turbulent flows but the model is not able to make real time future predictions given the historic data. \cite{raissi2017physics} applied a Galerkin finite element method with deep neural networks to solve PDEs automatically, what they call ``Physics-informed deep learning''.  Though these methods have shown the ability of deep learning in solving PDEs directly and deriving generalizable solutions, the key limitation of these approaches is that they require explicitly inputs of boundary conditions during inference, which are generally not available in real-time. \cite{Arvind} proposed a purely data-driven DL model for turbulence, compressed convolutional LSTM,  but the model lacks physical constraints and interpretability.  \cite{jinlong_2019} and \cite{constrain1} introduced statistical and physical constraints in the loss function to regularize the predictions of the model. However, their studies only focused on spatial modeling without temporal dynamics, besides regularization being ad-hoc and difficult to tune the hyper-parameters.  


\paragraph{\normalfont\textbf{Fluid Animation}} In parallel, the computer graphics community has also investigated using deep learning to speed up numerical simulations for generating realistic animations of fluids such as water and smoke. For example,  \cite{tompson2017accelerating} used an incompressible Euler's equation with a customized Convolutional Neural Network (CNN) to predict velocity update within a finite difference method solver. \cite{chu2017data} propose double CNN networks to synthesize high-resolution flow simulation based on reusable space-time regions. \cite{xie2018tempogan} and \cite{eulerian} developed deep learning models in the context of fluid flow animation, where physical consistency is less critical. \cite{Steffen} proposed a method for the data-driven inference of temporal evolutions of physical functions with deep learning. However, fluid animation emphases on the realism of the simulation rather than the physical consistency of the predictions or physics metrics and diagnostics of relevance to scientists. 

\paragraph{\normalfont\textbf{Video Prediction}} Our work is also related to future video prediction. Conditioning on the observed frames, video prediction models are trained
to predict future frames, e.g., \cite{mathieu2015deep, finn2016unsupervised, xue2016visual, villegas2017decomposing, Chelsea}. Many of these models are trained on natural videos with complex noisy data from unknown physical processes. Therefore, it is difficult to explicitly incorporate physical principles into the model. The turbulent flow problem studied in this work is substantially different from natural video prediction because it does not attempt to predict object or camera motions. Instead, our approach aims to emulate numerical simulations given noiseless observations from known governing equations. Hence, some of these techniques are perhaps under-suited for our application.

\section{Background in Turbulence Modeling}
Most fluid flows in nature are turbulent, but theoretical understanding of solutions to the governing equations, the Navier–Stokes equations, is incomplete. Turbulent fluctuations occur over a wide range of length and time scales with high correlations between these scales. Turbulent flows are characterized by chaotic motions and intermittency, which are difficult to predict.
\begin{figure}[hbt!]
\centering
\includegraphics[width= 0.45\textwidth]{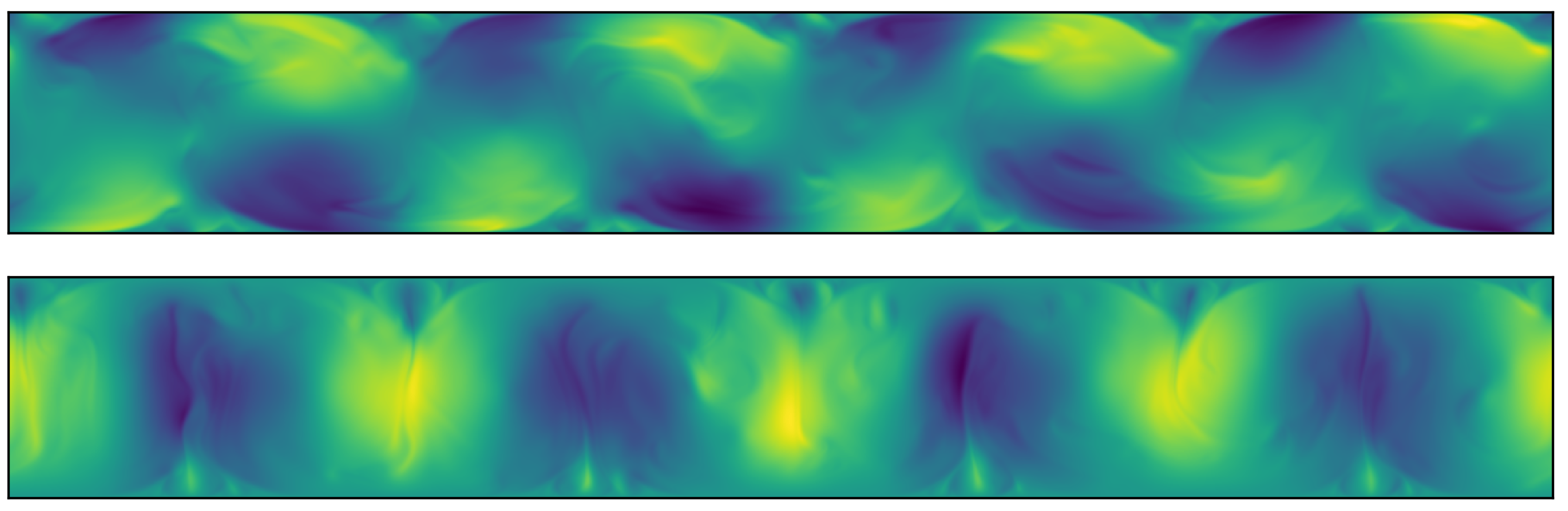}
\caption{A snapshot of the Rayleigh-B\'enard convection flow, the velocity fields along $x$ direction (top) and $y$ direction (bottom) \cite{Dragos-thesis}. The spatial resolution is 1792 x 256 pixels.}
\label{snapshot}
\end{figure}

The physical system we investigate is two-dimensional Rayleigh-B\'enard convection (RBC), a model for turbulent convection, with a horizontal layer of fluid heated from below so that the lower surface is at a higher temperature than the upper surface. Turbulent convection is a major feature of the dynamics of the oceans, the atmosphere, as well as engineering and industrial processes, which has motivated numerous experimental and theoretical studies for many years. The RBC system serves as an idealized model for turbulent convection that exhibits the full range of dynamics of turbulent convection for sufficiently large temperature gradients. 

Let $\bm{w}(t)$ be the vector velocity field of the flow with two components $(u(t),v(t))$, velocities along $x$ and $y$ directions, the governing equations for this physical system are:
\begin{align*}
&\nabla \cdot \bm{w} = 0 &\text{Continuity Equation} \nonumber\\ 
&\frac{\partial \bm{w}}{\partial t} + (\bm{w} \cdot \nabla) \bm{w}  = -\frac{1}{\rho_0} \nabla p + \nu \nabla^2 \bm{w} + f &\text{Momentum Equation} \nonumber\\ 
&\frac{\partial T}{\partial t} +  (\bm{w} \cdot \nabla) T = \kappa \nabla^2 T  &\text{Temperature Equation}
\label{eqn:navier-stokes}
\end{align*}
where $p$ and $T$ are pressure and temperature respectively, $\kappa$ is the coefficient of heat conductivity, $\rho_0$ is density at temperature at the beginning, $\alpha$ is the coefficient of thermal expansion, $\nu$ is the kinematic viscosity, $f$ the body force that is due to gravity. In this work, we use a particular approach to simulate RBC that uses a Boussinesq approximation, resulting in a divergence-free flow, so $\nabla \cdot \bm{w}$ should be zero everywhere \cite{Dragos-thesis}.  Figure \ref{snapshot} shows a snapshot in our RBC flow dataset.

CFD allows simulating complex turbulent flows, however, the wide range of scales makes it very challenging to accurately resolve all the scales. More precisely, fully resolving a complex turbulent flow numerically, known as direct numerical simulations (DNS), requires a very fine discretization of space-time, which makes the computation prohibitive even with advanced high-performance computing. Hence most CFD methods, like Reynolds-Averaged Navier-Stokes and Large Eddy Simulations (\cite{turb1, turb2, mcdonough2007introductory}, resort to resolving the large scales whilst modeling the small scales, using  various averaging techniques and/or low-pass filtering of the governing equations.
%
However, the unresolved processes and their interactions with the resolved scales are extremely challenging to model. CFD remains computationally expensive despite decades of advancements in turbulence modeling and HPC.

Deep learning (DL) is poised to accelerate and improve turbulent flow simulations because well-trained DL models can generate realistic instantaneous flow fields with physically accurate spatiotemporal coherence, without solving the complex nonlinear coupled PDEs that govern the system \cite{tompson2017accelerating, mazier1, mazier2}.
However, DL models are hard to train and are often used as "black boxes"  as they lack  knowledge of the underlying physics and are very hard to interpret. While these DL models may achieve low prediction errors they often lack scientific consistency and do not respect the physics of the systems they model. Therefore, it is critical to infusing known physics laws and design efficient turbulent flow prediction DL models that are not only accurate but also physically meaningful.

\section{Methodology}

\begin{figure}[t]
\includegraphics[width= 0.7\linewidth]{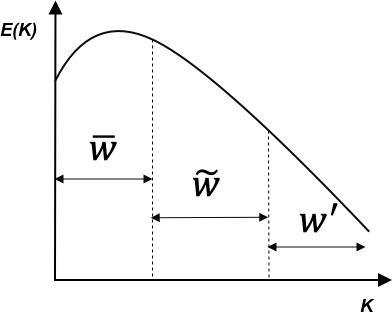}
\caption{Three level spectral decomposition of velocity $\bm{w}$, $E(k)$ is the energy spectrum and $k$ is wavenumber.}
\label{spec_decom}
\end{figure}

We aim to design physics-informed deep learning models by infusing CFD principles into deep neural networks.  The global idea behind our method is to  
decompose the turbulent flow into  components of different scales with trainable modules for simulating each component.
First, we provide a brief introduction of the CFD techniques which are built on this basic idea.

\subsection{Computational Fluid Dynamics}
Computational techniques are at the core of present-day turbulence investigations. Direct Numerical Simulation (DNS) are accurate but not computationally feasible for practical applications. Great emphasis was placed on the alternative approaches including Large-Eddy Simulation (LES)  and  Reynolds-averaged Navier–Stokes (RANS). See the book on turbulence \cite{turb1} for details. 

\begin{figure*}[!htb]
\includegraphics[width= 0.8\linewidth]{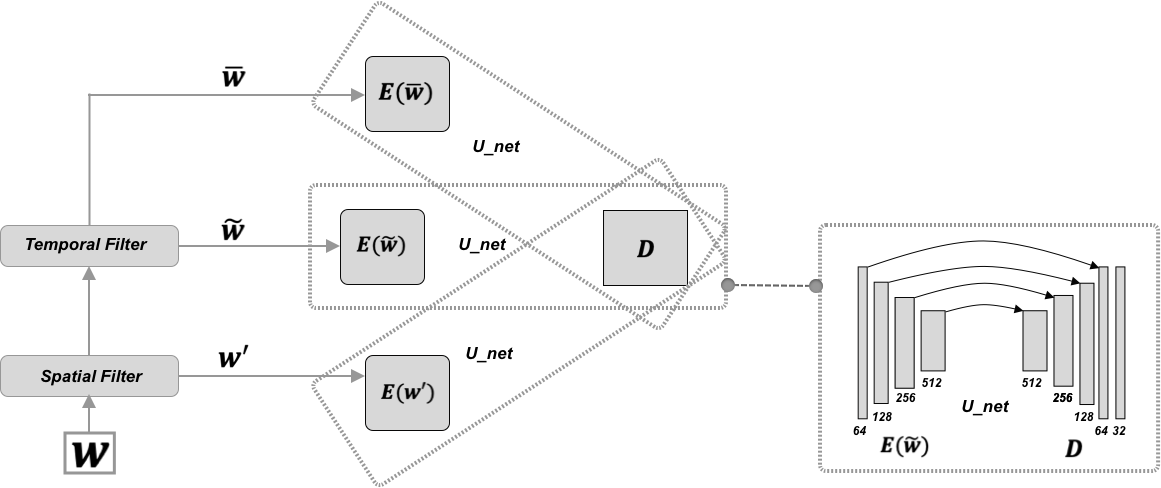}
    \caption{Turbulent Flow Net: three identical encoders to  learn the transformations of the three components of different scales, and one shared decoder that learns the interactions among these three components to generate the predicted  2D velocity field at the next instant. Each encoder-decoder pair can be viewed as a U-net and the aggregation is weighted summation.}
    \label{fig:model}
\end{figure*}

\textbf{Reynolds-averaged Navier–Stokes (RANS)} decomposes the turbulent flow $\bm{w}$ into two separable time scales: a time-averaged mean flow $\bar{\bm{w}}$ and a fluctuating quantity $\bm{w'}$. The resulting RANS equations contain a closure term, the Reynolds stresses, that require modeling, the classic closure problem of turbulence modeling. While this approach is a good first approximation to solving a turbulent flow, RANS does not account for broadband unsteadiness and intermittency, characteristic of most turbulent flows. Further, closure models for the unresolved scales are often inadequate, making RANS solutions to be less accurate. $n$ here is the moving average window size.
\begin{equation}
\bm{w}(\bm{x},t) = \bm{\bar{w}}(\bm{x},t) + \bm{w'} (\bm{x},t), \quad \bm{\bar{w}}(\bm{x},t) = \frac{1}{n}\int_{t-n}^t T(s)\bm{w}(\bm{x},s) ds
\end{equation}

\textbf{Large Eddy Simulation (LES)} is an alternative approach based on low-pass filtering of the Navier-Stokes equations that solves a part of the multi-scale turbulent flow corresponding to the most energetic scales. In LES, the large-scale component is a spatially filtered variable $\boldsymbol{\tilde{w}}$, which is usually expressed as a convolution product by the filter kernel $S$.  The kernel $S$ is often taken to be a Gaussian kernel. $\Omega_{i}$ is a subdomain of the solution and depends on the filter size \cite{LESbook}. 
\begin{equation}
\bm{w}(\bm{x},t) = \bm{\tilde{w}}(\bm{x},t) + \bm{w'} (\bm{x},t), \quad  \bm{\tilde{w}}(\bm{x},t) = \int_{\Omega_{i}} S(\bm{x}|\bm{\xi})\bm{w}(\bm{\xi},t) d\bm{\xi}
\end{equation}
The key difference between RANS and LES is that RANS is based on time averaging, leading to simpler steady equations, whereas LES is based on a spatial filtering process which is more accurate but also computationally more expensive.

\textbf{Hybrid RANS-LES Coupling} combines both RANS and LES approaches in order to take advantage of both methods \cite{turb3, hybrid1}. It decomposes the flow variables into three parts: mean flow, resolved fluctuations and unresolved (subgrid) fluctuations. RANS-LES coupling applies the spatial filtering operator $S$ and the temporal average operator $T$ sequentially. We can define $\bm{\bar{w}}$ in discrete form with using $\bm{w^*}$ as an intermediate term, 
\begin{align}
&\bm{w^*(\bm{x},t)} = S \ast\bm{w}  = \sum_{\bm{\xi}} S(\bm{x}|\bm{\xi})\bm{w}(\bm{\xi},t) \\
&\bm{\bar{w}(\bm{x},t)} = T \ast \bm{w^*} = \frac{1}{n}\sum_{s = t-n}^t  T(s)\bm{w^*}(\bm{x},s) 
\end{align}
then $\bm{\tilde{w}}$ can be defined as the difference between $\bm{w^*}$ and $\bm{\bar{w}}$:
\begin{align}
&\bm{\tilde{w}} = \bm{w^*} - \bm{\bar{w}}, \quad \bm{w'} = \bm{w} - \bm{w^{*}}
\end{align}
Finally we can have the three-level decomposition of the velocity field.
\begin{equation}
\bm{w} = \bm{\bar{w}} + \bm{\tilde{w}} + \bm{w^{'}}
\end{equation}
Figure \ref{spec_decom} shows this three-level decomposition in wavenumber space \cite{turb3}. $k$ is the wavenumber, the spatial frequency in the Fourier domain. $E(k)$ is the energy spectrum describing how much kinetic energy is contained in eddies with wavenumber $k$. Small $k$ corresponds to large eddies that contain most of the energy. The slope of the spectrum is negative and indicates the transfer of energy from large scales of motion to the small scales. This hybrid approach combines computational efficiency of RANS with the resolving power of LES to provide a technique that is less expensive and more tractable than pure LES. 

\subsection{Turbulent Flow Net}
Inspired by techniques used in hybrid RANS-LES Coupling  to separate scales of a multi-scale system, we propose  a hybrid deep learning framework, \tf{},   based on the multi-level spectral decomposition of the turbulent flow. 

Specifically, we decompose the velocity field into three components of different scales using two scale separation operators, the spatial filter $S$ and the temporal filter $T$. In traditional CFD, these  filters are usually pre-defined, such as the Gaussian spatial filter. In  \tf{}, both filters are \textit{trainable} neural networks. The spatial filtering process is instantiated as one layer convolutional neural network  with a single 5$\times$5 filter to each input image. The temporal filter is also implemented as a convolutional layer with a single 1$\times$1 filter applied to every $T$ images. The motivation for this design is to explicitly guide the DL model to learn the non-linear dynamics of both large and small eddies as relevant to the task of spatio-temporal prediction. 

We design three identical encoders to encode the three scale components separately. We use a shared decoder to learn the interactions among these three components and generate the final prediction. Each encoder and the decoder can be viewed as a U-net without duplicate layers and middle layer in the original architecture \cite{unet}. The encoder consists of four convolutional layers with double the number of feature channels of the previous layer and stride 2 for down-sampling. The decoder consists of one output layer and four deconvolutional layers with summation of the corresponding feature channels from the three encoders and the output of the previous layer as input. Figure \ref{fig:model} shows the overall architecture of our hybrid model \tf{}. 

To generate multi-step forecasts, we perform one-step ahead prediction and roll out the predictions autoregressively.  Furthermore, we also consider a variant of \tf{} by explicitly adding physical constraint to the loss function. Since the turbulent flow under investigation has zero divergence ($\nabla \cdot \bm{w}$ should be zero everywhere), we  include $||\nabla \cdot \bm{w}||^2$ as a regularizer to constrain the predictions, leading to a constrained \tf{}, or \texttt{Con TF-Net}.

\section{Experiments}
\begin{table*}
\ra{1.3}
\normalsize
\centering
\begin{center}
\begin{tabular}{|p{3.8cm}|p{1.6cm}|p{1.6cm}|p{1.6cm}|p{1.6cm}|p{1.6cm}|p{1.6cm}|p{1.6cm}|}
\hline
\textbf{Models} & \textbf{TF-net} & \textbf{U\_net} & \textbf{GAN} & \textbf{ResNet} & \textbf{ConvLSTM} & \textbf{SST} & \textbf{DHPM} \\ \hline
\textbf{\#params(10\textasciicircum{}6)} & 15.9 & 25.0 & 26.1 & 21.2 & 11.8 & 49.9 & 2.12 \\ \hline
\textbf{input length} & 25 & 25 & 24 & 26 & 27 & 23 & \textbackslash{} \\ \hline
\textbf{\#accumulated errors} & 4 & 6 & 5 & 5 & 4 & 5 & \textbackslash{} \\ \hline
\textbf{time for one epoch(min)} & 0.39 & 0.57 & 0.73 & 1.68 & 45.6 & 0.95 & 4.591 \\ \hline
\end{tabular}%
\end{center}
\caption{The number of parameters, the best number of input frames, the best number of accumulated errors for backpropogation and training time for one epoch on 8 V100 GPUs for each model.}
\label{table}
\vspace{-5mm}
\end{table*}

\subsection{Dataset}
The dataset for our experiments comes from two dimensional turbulent flow simulated using the Lattice Boltzmann Method \cite{Dragos-thesis}. We use only the velocity vector fields, where the spatial resolution of each image (snapshots in time) is 1792 x 256. Each image has two channels, one is the turbulent flow velocity along $x$ direction and the other one is the velocity along $y$ direction. The physics parameters relevant to this numerical simulation are: Prandtl number $=0.71$, Rayleigh number $=2.5\times 10^8$ and the maximum Mach number $=$ 0.1. We use 1500 images  for our experiments. The task is to predict the spatiotemporal velocity fields up to $60$ steps ahead given $10$ initial frames. 
\begin{figure*}[hbt!]
  \begin{minipage}[b]{0.32\textwidth}
    \includegraphics[width=\textwidth]{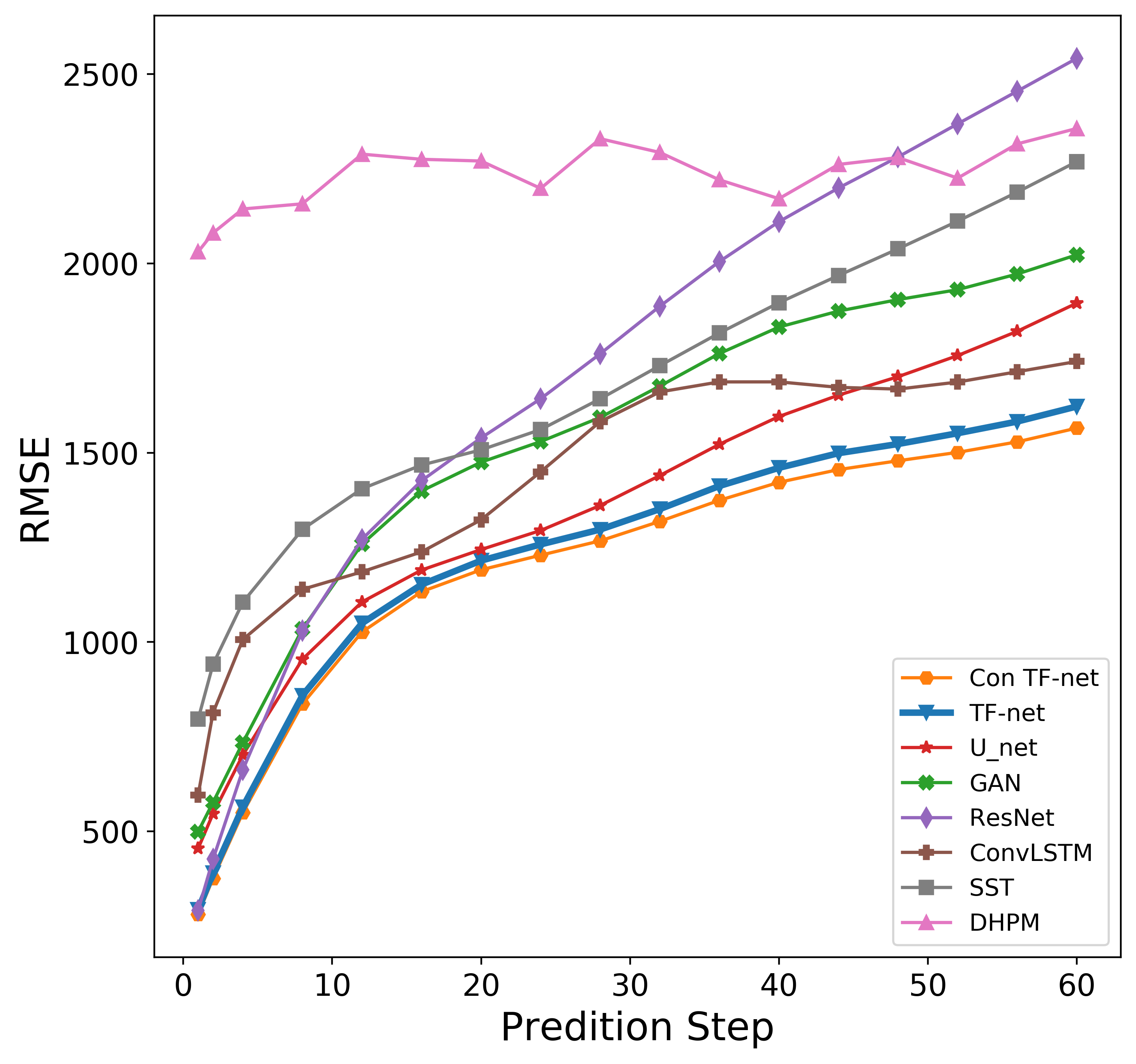}
    \caption{Root mean square errors of different models' predictions at varying forecasting horizon}
    \label{rmse}
  \end{minipage} \hfill
  \begin{minipage}[b]{0.32\textwidth}
     \includegraphics[width=\textwidth]{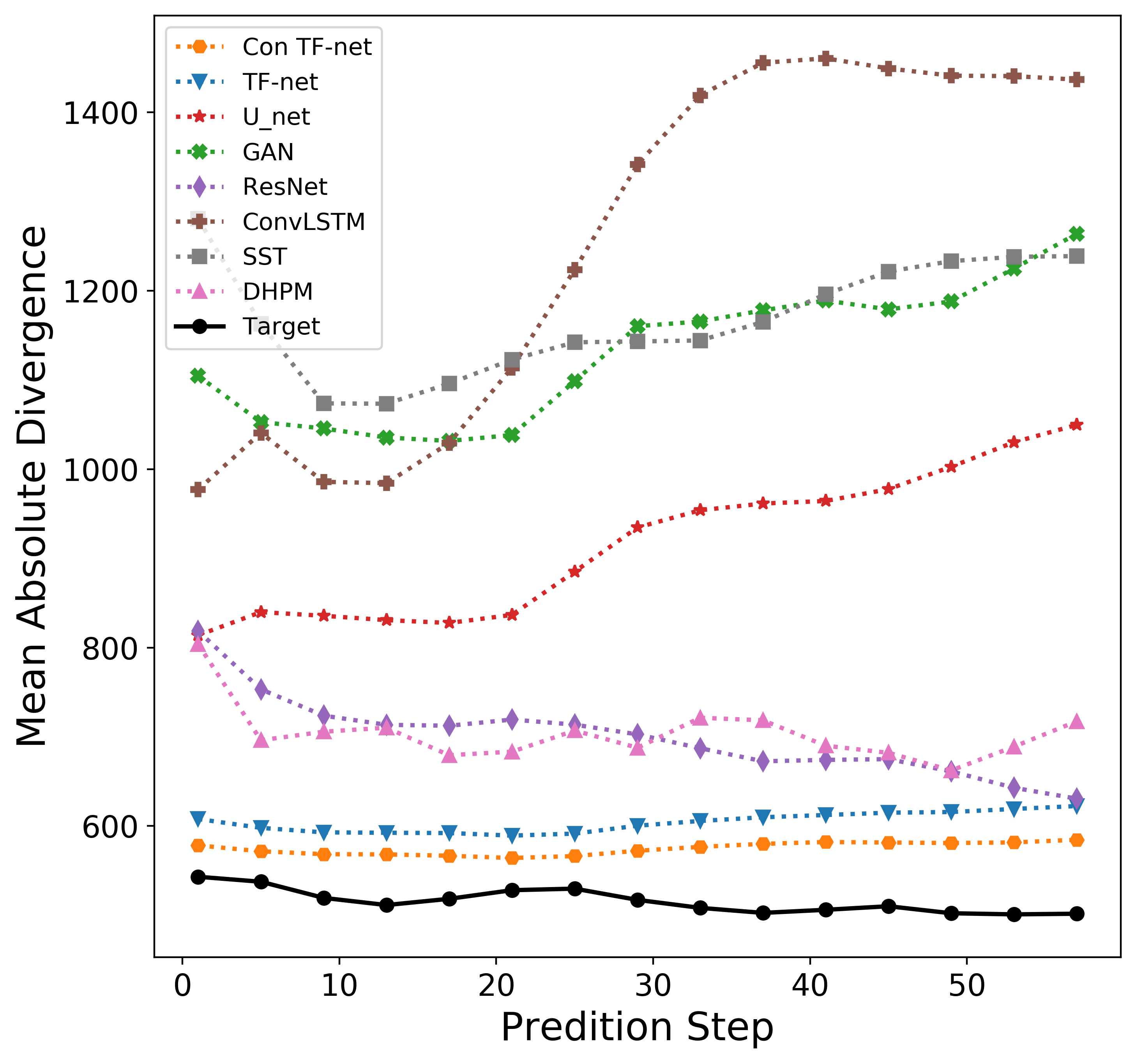}
    \caption{Mean absolute divergence of different models' predictions at varying forecasting horizon}
    \label{div}
  \end{minipage} \hfill
  \begin{minipage}[b]{0.316\textwidth}
\includegraphics[width=\linewidth]{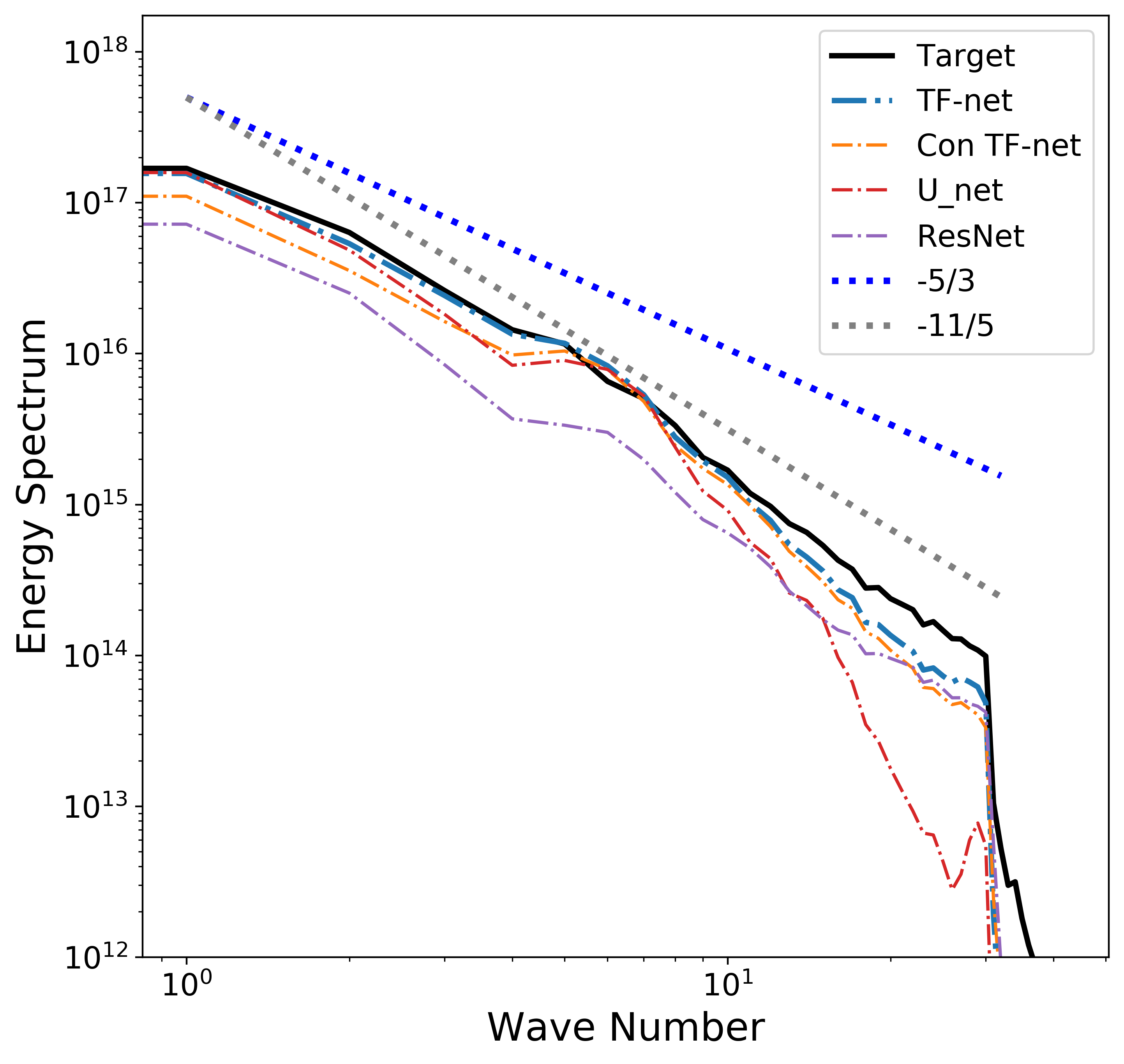}
\caption{The Energy Spectrum of \tf{}, \texttt{U-net} and \texttt{ResNet} on the leftmost square sub-region. }
\label{spectrum}
  \end{minipage} 
\end{figure*}

\begin{figure*}[hbt!]
\centering\includegraphics[width= 0.98\textwidth]{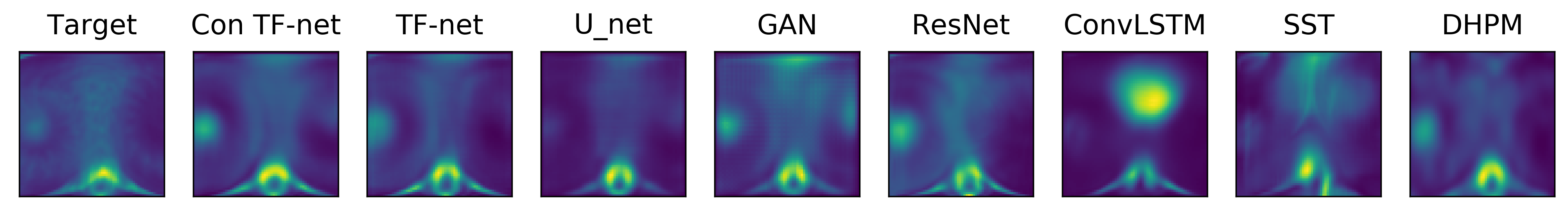}
\caption{Turbulence kinetic energy of all models' predictions at the leftmost square field in the original rectangular field }
\label{tke}
\end{figure*} 

\begin{figure}[htb!]
\includegraphics[width= 0.95\linewidth]{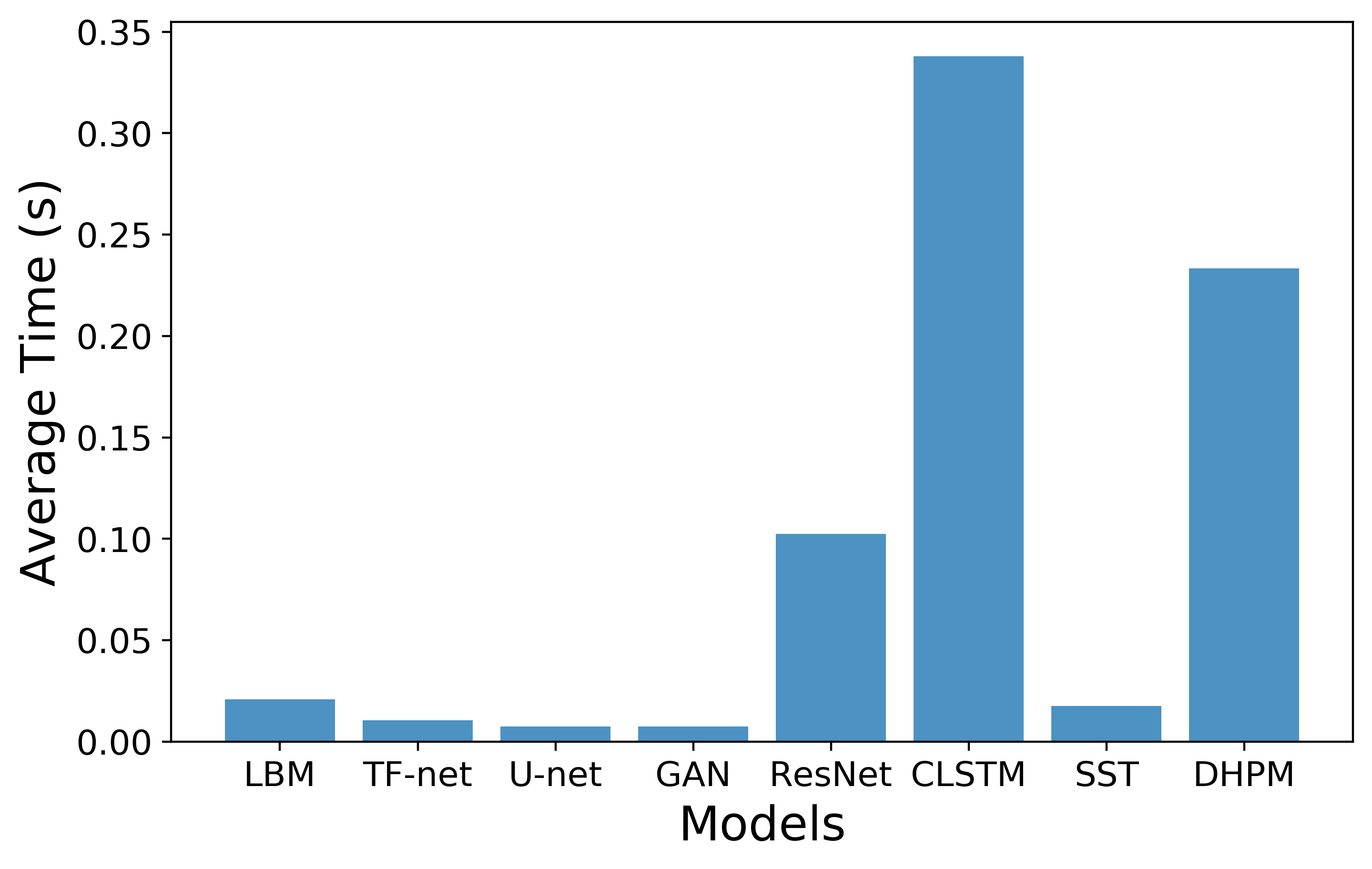}
\hspace*{0cm}
\caption{Average time to produce one 64 $\times$ 448 2D velocity field for all models on single V100 GPU.}
\label{time}
\end{figure}

We divided each 1792 by 256 image into 7 square sub-regions of size 256 x 256, then downsample them into 64 x 64 pixels sized images. We use a sliding window approach to generate 9,870 samples of sequences of velocity fields: 6,000 training samples, 1,700 validation samples and 2,170 test samples. The DL model is trained using back-propagation through prediction errors accumulated over multiple steps. We use a validation set for hyper-parameters tuning based on the average error of predictions up to six steps ahead. The hyper-parameters tuning range can be found in Table \ref{range} in the appendix. All results are averaged over three runs with random initialization. 

\begin{figure*}[htb!]
\centering
\begin{subfigure}
\centering
\includegraphics[width=0.49\textwidth]{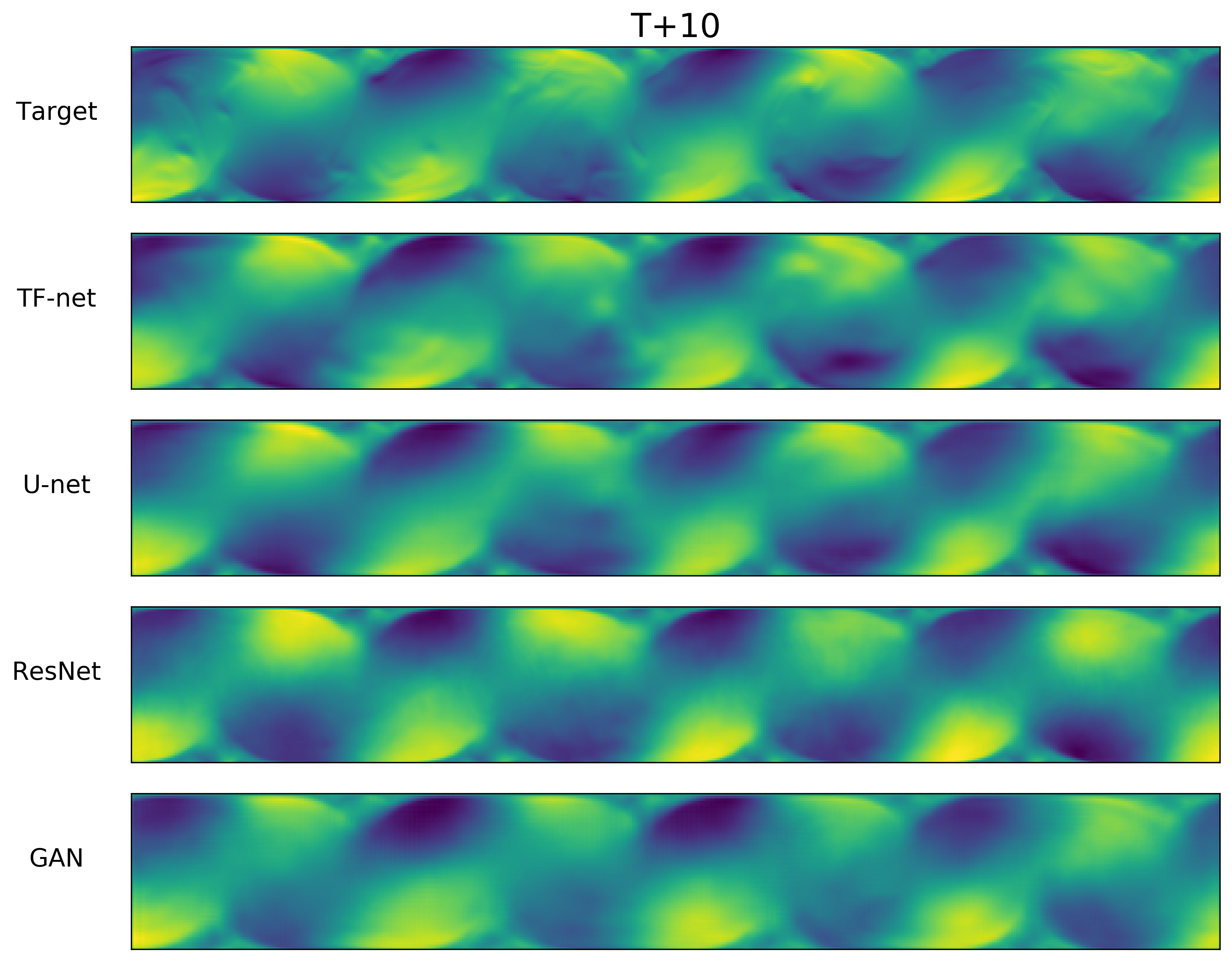}
\end{subfigure}
\hfill
\begin{subfigure} 
\centering 
\includegraphics[width=0.49\textwidth]{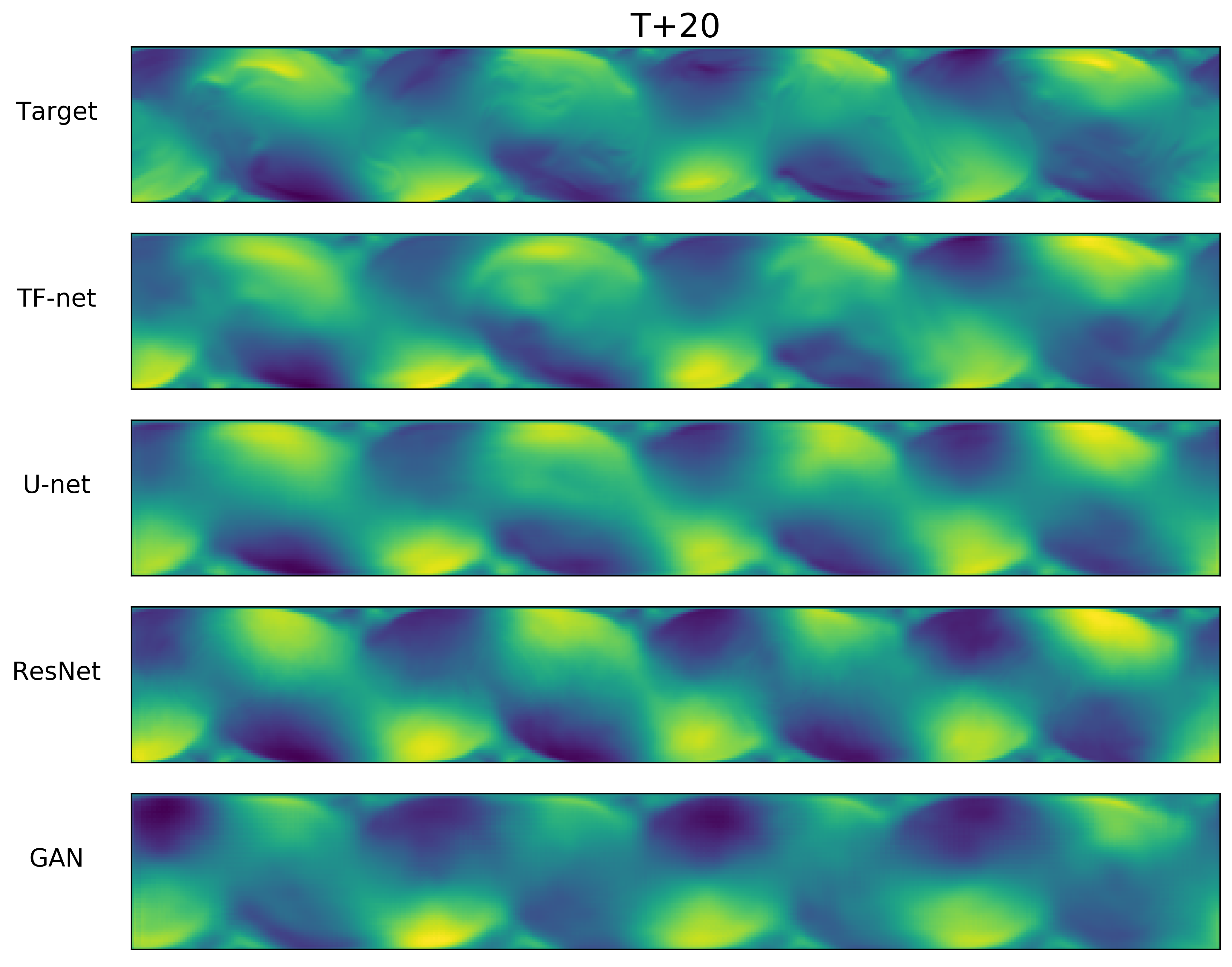}
\end{subfigure}
\begin{subfigure} 
\centering 
\includegraphics[width=0.49\textwidth]{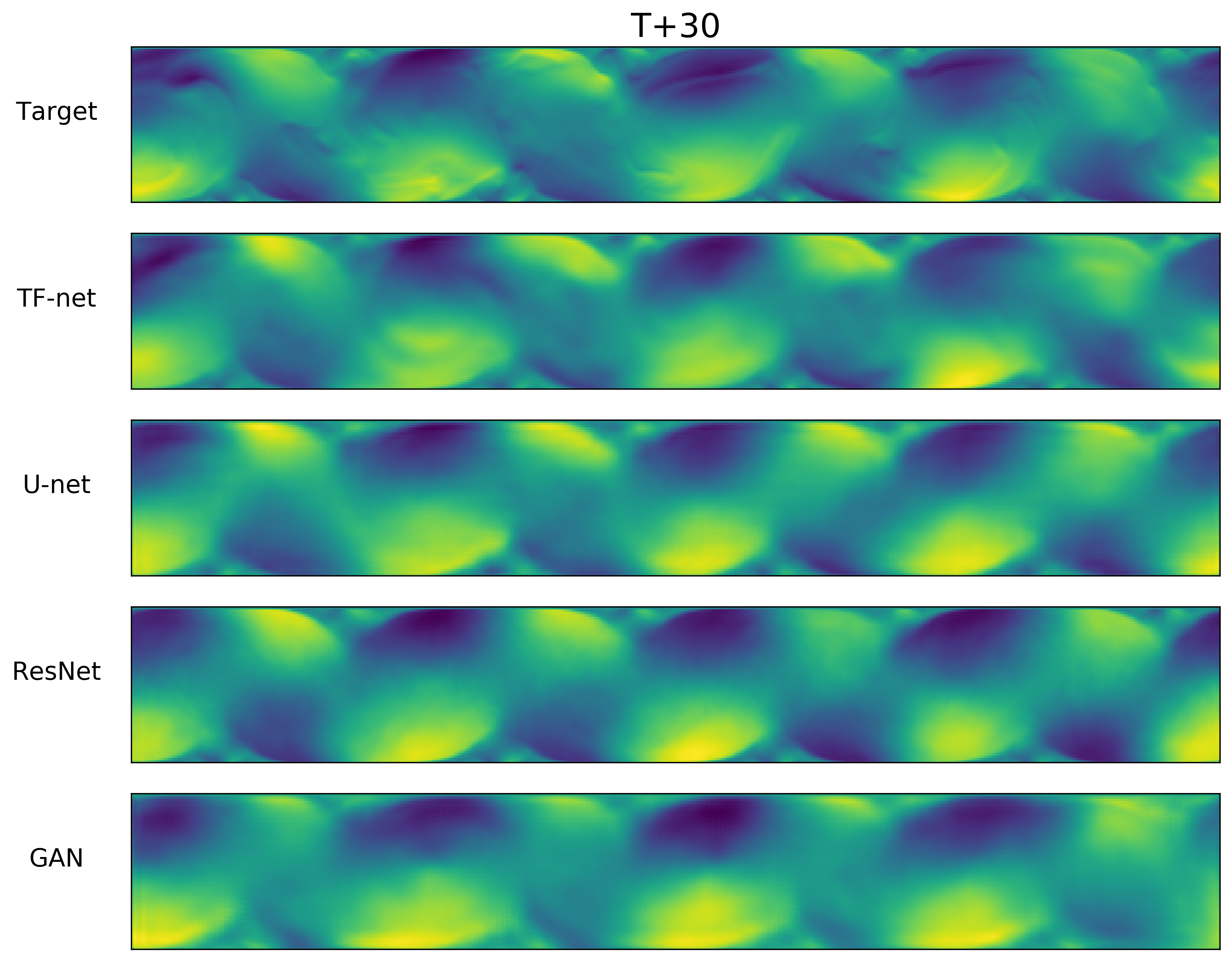}
\end{subfigure}
\hfill
\begin{subfigure}  
\centering 
\includegraphics[width=0.49\textwidth]{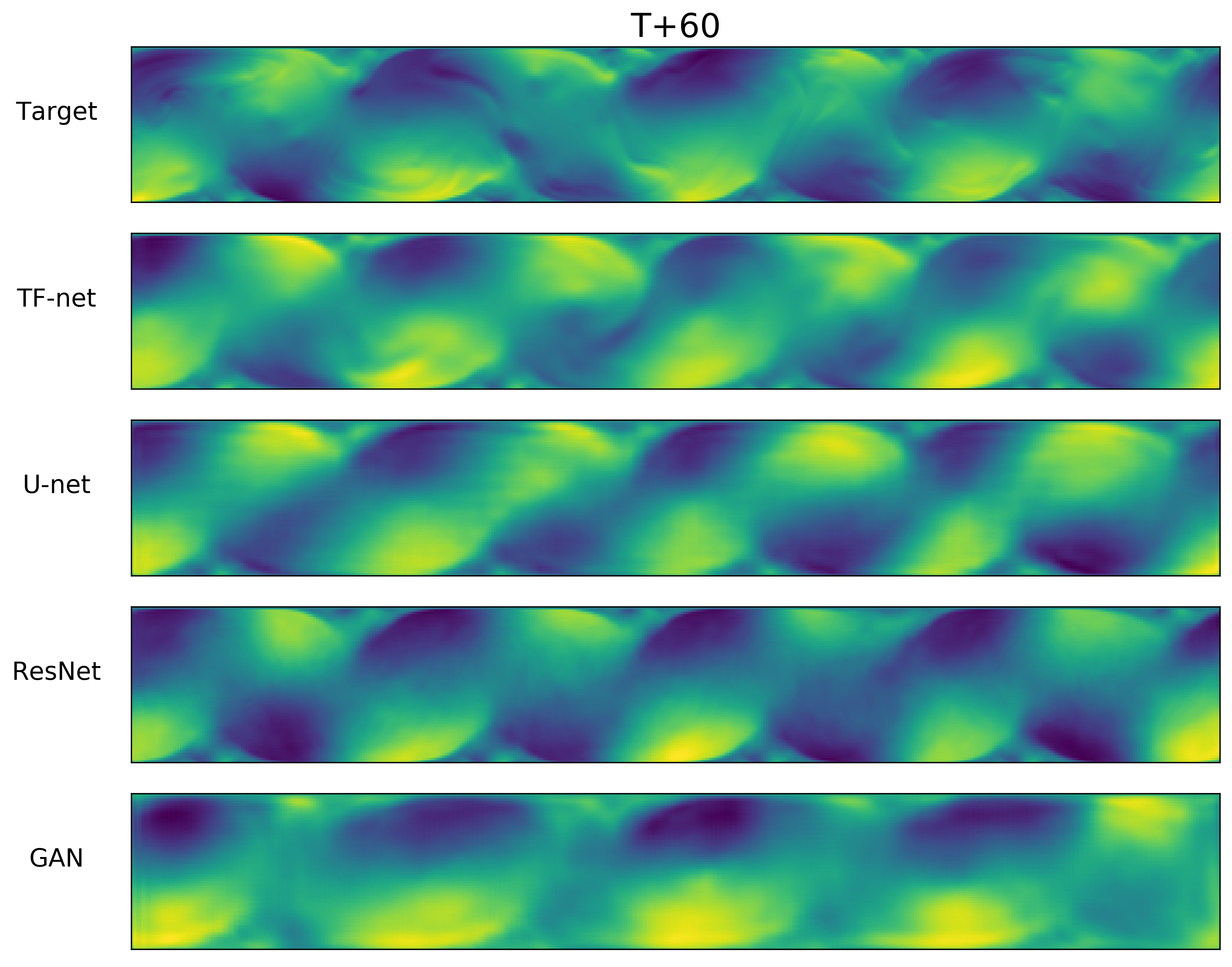}
\end{subfigure}
\caption{Ground truth (target) and predicted $u$ velocities by \tf{} and three best baselines (U-Net, ResNet and GAN) at time $T+10$, $T+20$, $T+30$ to $T+60$ (suppose $T$ is the time step of the last input frame).}
\label{vel}
\end{figure*}

\subsection{Baseline}
We compare our model with a series of state-of-the-art  baselines for turbulent flow prediction. 
\begin{enumerate}
\item \texttt{ResNet} \cite{He2015}: a 34-layer residual convolutional neural networks by replacing the final dense layer with a convolutional layer with two output channels.
\vspace{0.3cm}
\item \texttt{ConvLSTM} \cite{convlstm}: a 3-layer Convolutional LSTM model used for spatiotemporal precipitation nowcasting.
\vspace{0.3cm}
\item \texttt{U-Net} \cite{unet}: Convolutional neural networks developed for image segmentation, also used for video prediction.
\vspace{0.3cm}
\item \texttt{GAN}: \texttt{U-Net} trained with a convolutional discriminator.
\vspace{0.3cm}
\item \texttt{SST} \cite{PDE-CDNN}:  hybrid physics-guided deep learning model using warping scheme for linear energy equation to predict sea surface temperature, which is also applicable to the linearized momentum equation that governs the velocity fields. 
\vspace{0.3cm}
\item \texttt{DHPM} \cite{dhpm}: Deep Hidden Physics Model is to  directly approximate the solution of partial differential equations with fully connected networks using space time location as inputs. The model is trained twice on the training set and  the test set with boundary conditions. This model can be formulated as, $\textsl{Loss = } ||\textbf{w} - \hat{\textbf{w}}|| +  ||\nabla \cdot \hat{\textbf{w}}|| + ||\hat{\textbf{w}_t} + (\hat{\textbf{w}} \cdot \nabla)\hat{\textbf{w}} - \nu\nabla^2 \hat{\textbf{w}} - f||$, where
$\hat{\textbf{w}} = \textsl{NN}(x, y, t) $, and $f = \textsl{NN}(x, y, t, \hat{u}, \hat{v}, \hat{u}_x, \hat{v}_x, \hat{u}_y, \hat{v}_y)$.
\vspace{0.3cm}
\end{enumerate}
Here \texttt{ResNet}, \texttt{ConvLSTM}, \texttt{U-net} and \texttt{GAN} are pure data-driven spatiotemporal deep learning models for video predictions. \texttt{SST} and \texttt{DHPM} are hybrid physics-informed deep learning that aim to incorporate prior physical knowledge into deep learning for fluid simulation.

\subsection{Evaluation Metrics}
Even though Root Mean Square Error (RMSE) is a widely accepted metric for quantifying the prediction performance, it only measures pixel differences. We need to check whether the predictions are physically meaningful and preserve desired physical quantities, such as Turbulence Kinetic Energy, Divergence and Energy Spectrum. Therefore, we include a set of additional metrics for evaluation. 

\textbf{Root Mean Square Error}
We calculate the RMSE of all predicted values from the ground truth for each pixel. 

\textbf{Divergence} Since we investigate incompressible turbulent flows in this work, which means the divergence, $\nabla \cdot {\textbf{w}}$, at each pixel should be zero, we use the average of absolute divergence over all pixels at each prediction step as an additional evaluation metric. 

\textbf{Turbulence Kinetic Energy}
In fluid dynamics, turbulence kinetic energy is the mean kinetic energy per unit mass associated with eddies in turbulent flow. Physically, the turbulence kinetic energy is characterised by measured root mean square velocity fluctuations, 
\begin{equation}
(\overline{(u^{'})^2} + \overline{(v^{'})^2})/2, \quad  \overline{(u^{'})^2} = \frac{1}{T}\sum_{t=0}^T(u(t) - \bar{u})^2
\end{equation}
where $t$ is the time step. We calculate the turbulence kinetic energy for each predicted sample of 60 velocity fields.

\textbf{Energy Spectrum}
The energy spectrum of turbulence, $E(k)$, is related to the mean turbulence kinetic energy as 
\begin{equation}
\int_{0}^{\infty}E(k)dk = (\overline{(u^{'})^2} + \overline{(v^{'})^2}),
\end{equation}
where $k$ is the wavenumber, the spatial frequency in 2D Fourier domain. We calculate the Energy Spectrum on the Fourier transformation of the Turbulence Kinetic Energy fields. The large eddies have low wavenumbers and the small eddies correspond to high wavenumbers. The spectrum indicates how much kinetic energy is contained in eddies with wavenumber $k$.


\begin{figure}[ht]
\includegraphics[width= \linewidth]{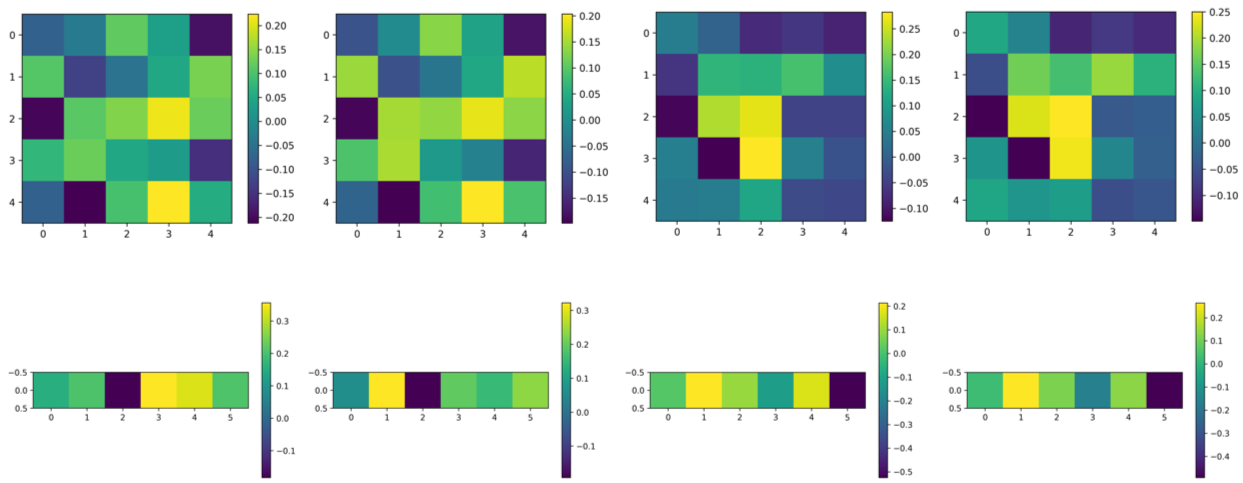}
\caption{Learned spatial and temporal filters in \tf{}}
\label{filters}
\vspace{-5mm}
\end{figure}

\begin{figure}[htb!]
\includegraphics[width= 0.95\linewidth]{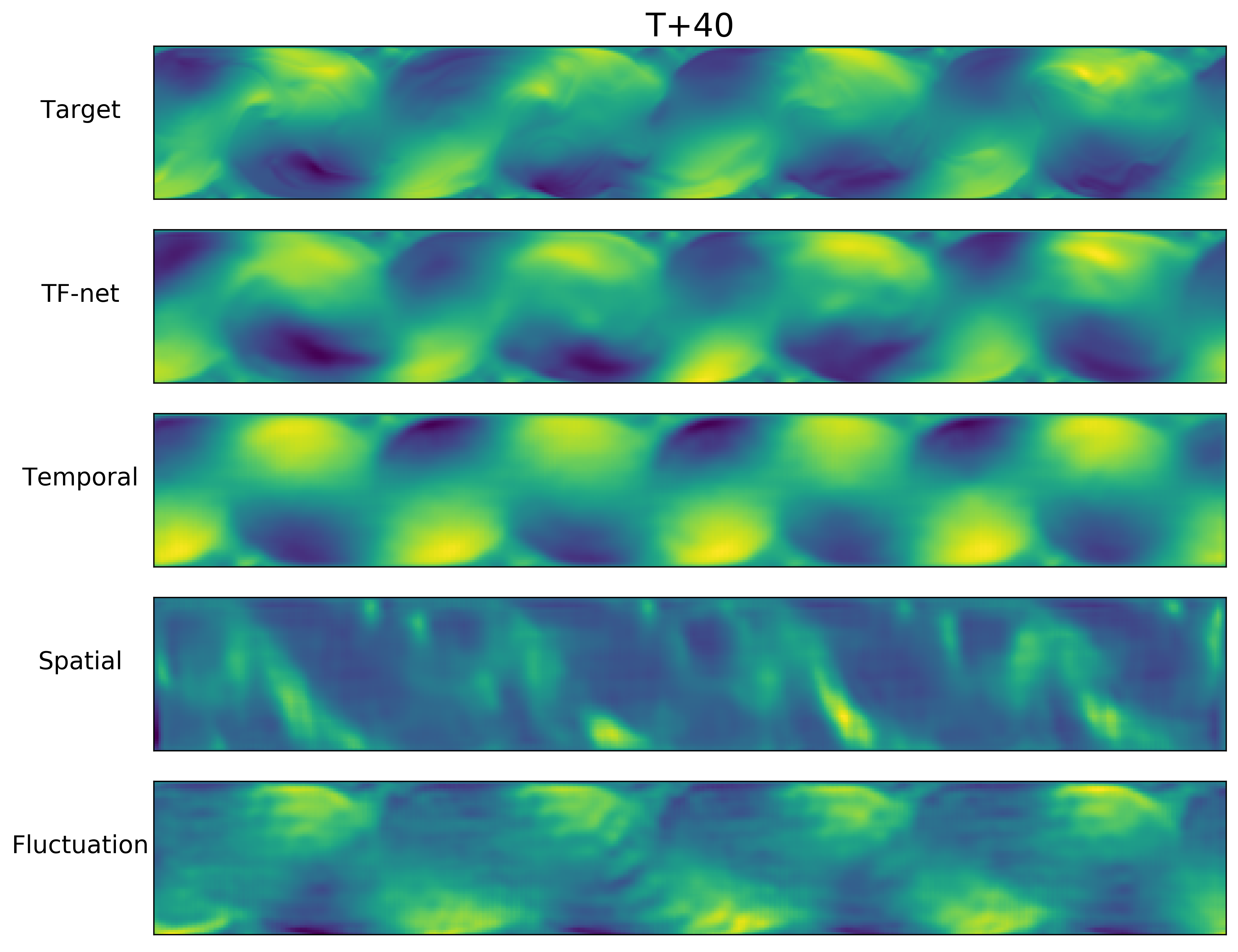}
\hspace*{0cm}
\caption{Ablation study: From top to bottom are the target, the predictions of \tf{}, and the outputs of each small U-net of \tf{} while the other two encoders are zeroed out at time $T+40$}
\label{ablation}
\vspace{-5mm}
\end{figure}





\subsection{Accuracy and Efficiency}
Figure \ref{rmse} shows the growth of RMSE with prediction horizon up to $60$ time steps ahead. \tf{} consistently outperforms all baselines, and constraining it with divergence free regularizer can  further improve the performance. We also found \texttt{DHPM} is able to overfit the training set but performs poorly when tested outside of the training domain. Neither Dropout nor regularization techniques can improve its performance. Also, the warping scheme of the \cite{PDE-CDNN} relies on the simplified linear assumption, which was too limiting for our non-linear problem.

Figure \ref{div} shows the averages of absolute divergence over all pixels at each prediction step. \tf{} has lower divergence than other models even without additional divergence free constraint for varying prediction step. It is worth mentioning that there is a subtle trade-off between RMSE and divergence. Even though constraining model with the divergence-free regularizer can reduce the divergence of the model predictions, too much constraint also has the side effect of smoothing out the small eddies, which results in a larger RMSE. 

Table \ref{table} displays the number of parameters, the best number of input frames, the best number of accumulated errors for backpropogation and training time for one epoch on 8 V100 GPUs for each model. Our model has significantly smaller number of parameters than most baselines yet achieves the best performance. About 25 historic images are enough for our model to generate reasonable predictions, and ConvLSTM require large memory and training time, especially when the number of historic input frames is large. Additionally, Table \ref{range} in appendix displays the hyper-parameters tuning range. 

Figure \ref{time} shows the average time to produce one 64 $\times$ 448 2d velocity field for all models on single V100 GPU. We can see that \texttt{TF-net}, \texttt{U\_net} and \texttt{GAN} are faster than the numerical method. The reason why speed up is not significant is that the numerical model is highly optimized at the bit level and the LBM method used to generate these fields is already an highly optimized approximation to the Navier Stokes equations that may be difficult/unfair to beat from a runtime perspective. We believe that \tf{} will show greater advantage of speed on higher resolution data, and unlike the numerical method, it can generalize to different datasets.

\subsection{Prediction Visualization}
Figure \ref{tke} displays predicted TKEs of all models at the leftmost square field in the original rectangular field.  Figure \ref{spectrum} shows the energy spectrum of our model and two best baseline at the leftmost square sub-field. While the turbulence kinetic energy of  \tf{}, \texttt{U-net} and \texttt{ResNet} appear to be similar in Figure \ref{tke}, from the energy spectrum in Figure \ref{spectrum}, we can see that \tf{} predictions are in fact much closer to the target. Extra divergence free constraint does not affect the energy spectrum of predictions. By Incorporating physics principles in deep learning, \tf{} is able to generate predictions that are physically consistent with the ground truth. 

\begin{figure*}[htb!]
\centering
\begin{subfigure}
\centering
\includegraphics[width=0.24\textwidth]{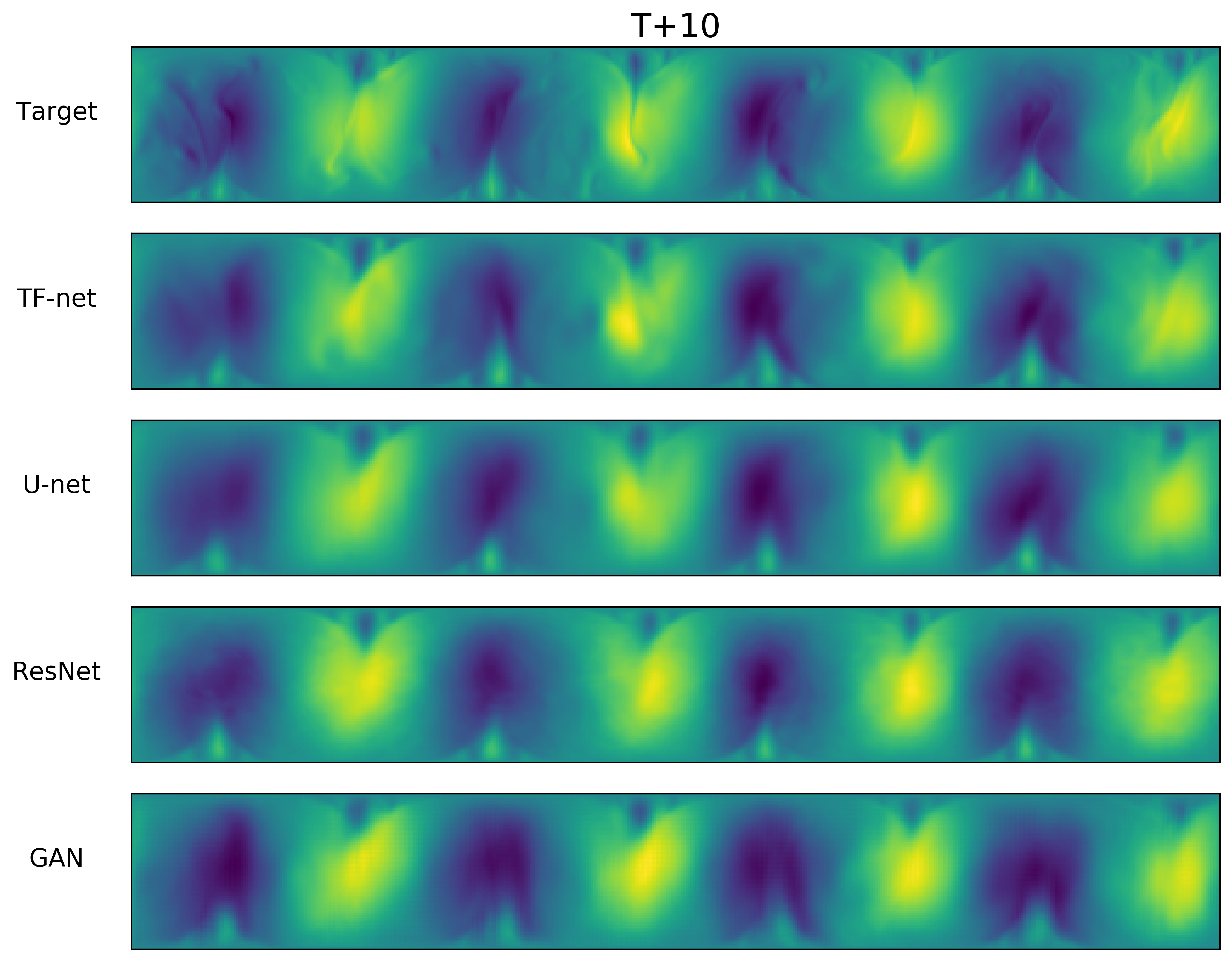}
\end{subfigure}
\begin{subfigure} 
\centering 
\includegraphics[width=0.24\textwidth]{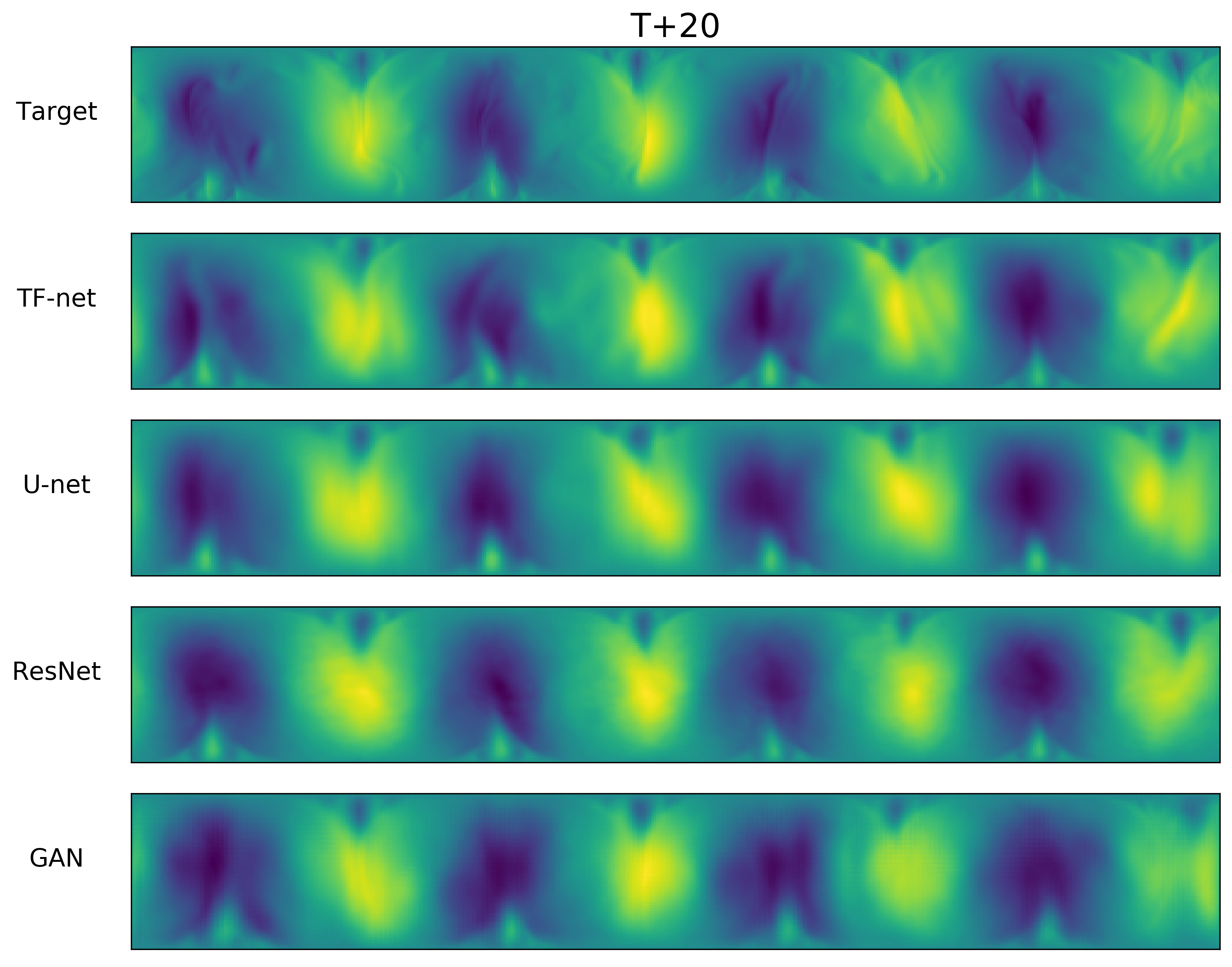}
\end{subfigure}
\begin{subfigure} 
\centering 
\includegraphics[width=0.24\textwidth]{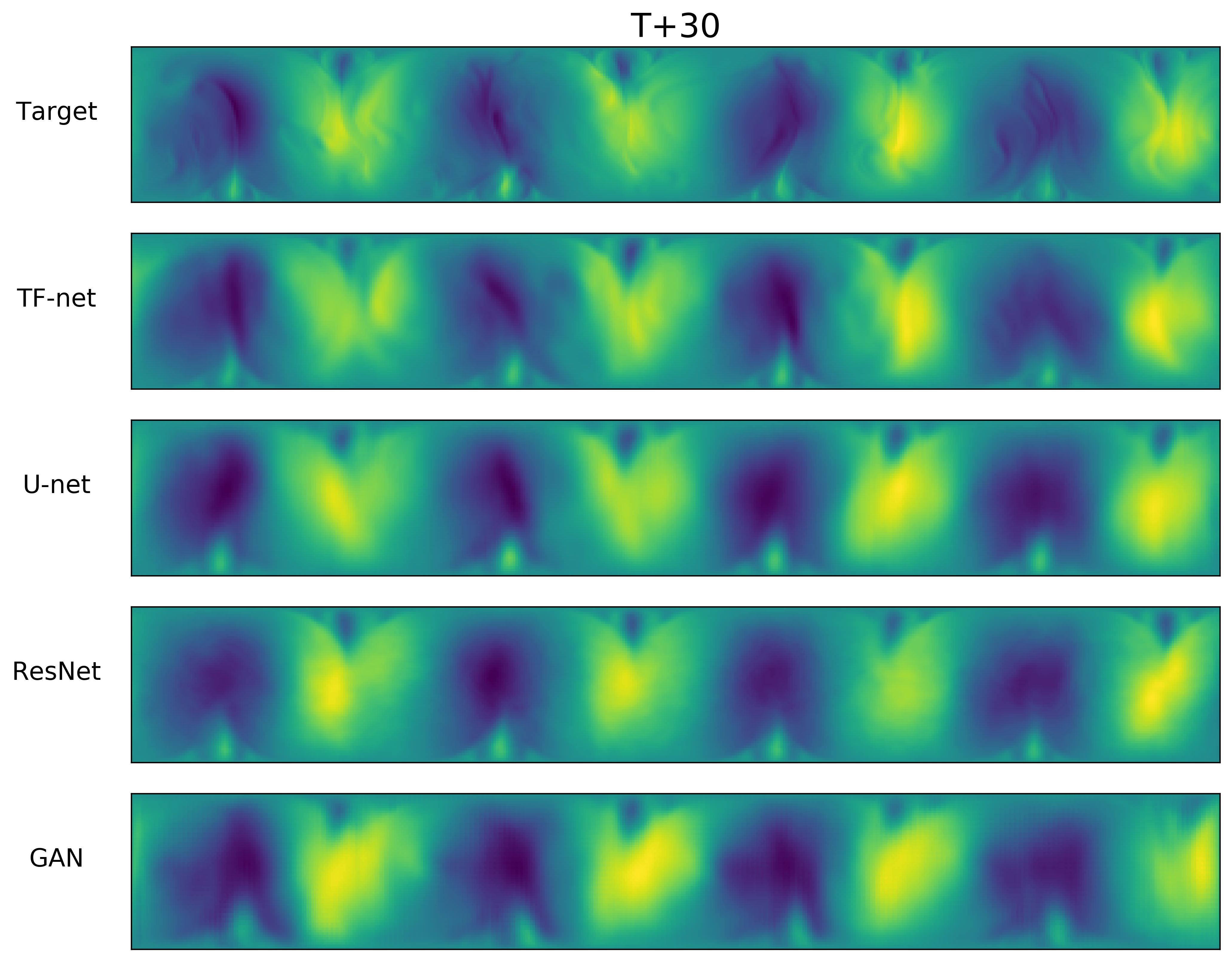}
\end{subfigure}
\begin{subfigure}  
\centering 
\includegraphics[width=0.24\textwidth]{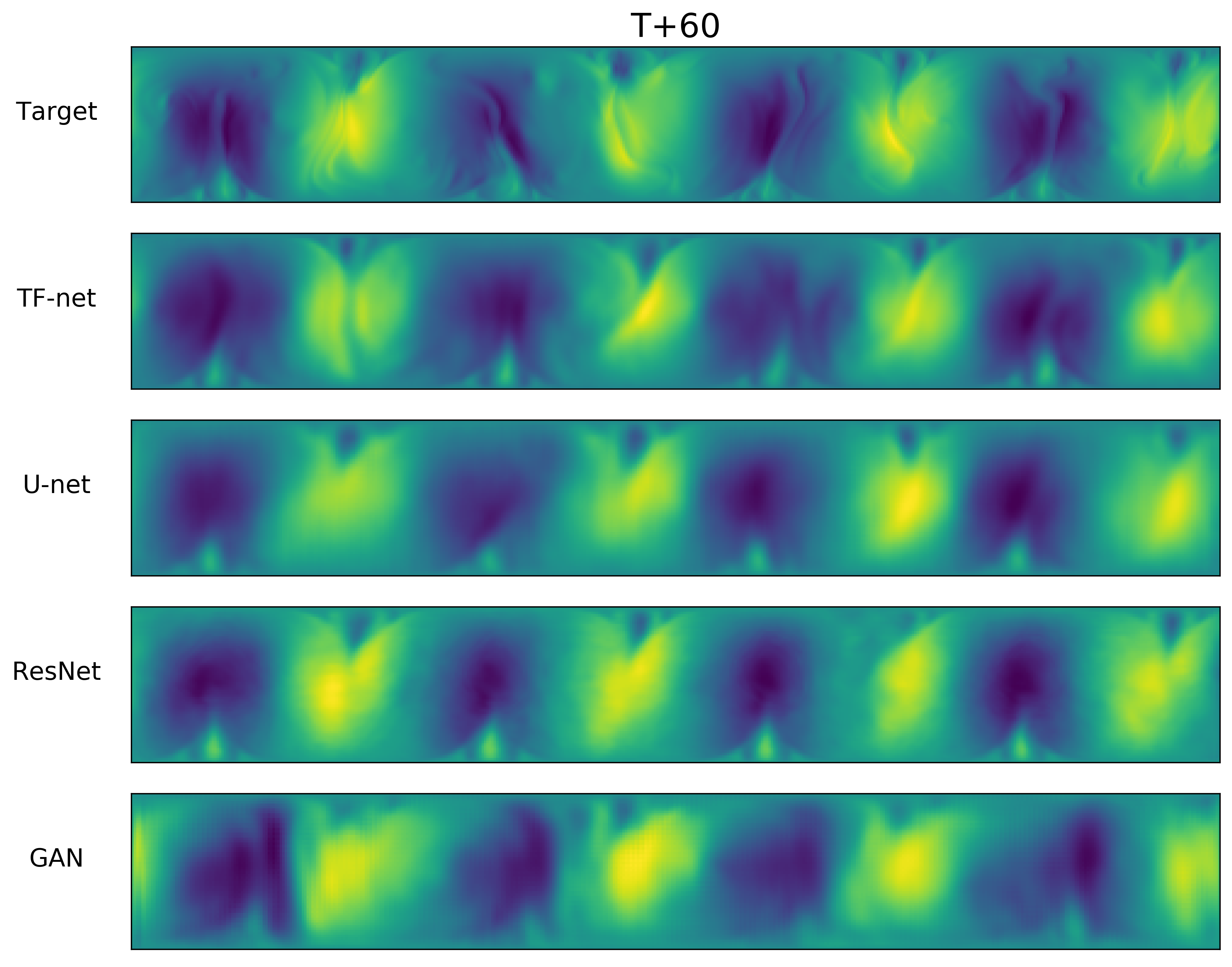}
\end{subfigure}
\caption{Ground truth and predicted $v$ velocities by models, suppose $T$ is the time step of the last input image.(suppose $T$ is the time step of the last input frame).}
\label{vel_v}
\end{figure*}

During Inference, we apply the trained \tf{} to the entire input domain instead of square sub-regions. Figure \ref{vel} and Figure \ref{vel_v} shows the ground truth and the predicted $u$ and $v$ velocity fields from all models from time step $0$ to $60$. We also provide videos of predictions by \tf{} and several best baselines in \url{https://www.youtube.com/watch?v=80U8lcIZYe4} and \url{https://www.youtube.com/watch?v=7J0RNiou5-4}, respectively. We see that the predictions by our \tf{} model are the closest to the target based on the shape and the frequency of the motions. Baselines generate smooth predictions and miss the details of small scale motion. \texttt{U-net} is the best performing data-driven video prediction baseline. \cite{marie} also found the U-net architecture is quite effective in modeling dynamics flows. Nevertheless, there is still room for improvement in long-term prediction for all the models.

\subsection{Ablation Study}
We also perform an additional ablation study of \tf{} to understand each component of \tf{} and investigate whether the \tf{} has actually learned the flow with different scales. During inference, we applied each small U-net in \tf{} with the other two encoders removed to the entire input domain. Figure \ref{ablation} (The full video can be found on \url{https://www.youtube.com/watch?v=ysdrMUfdhe0}) includes the predictions of \tf{}, and the outputs of each small U-net while the other two encoders are zeroed out at $T+40$. We observe that the outputs of each small u-net are the flow with different scales, which demonstrates that \tf{} can learn multi-scale behaviors of turbulent flows.  We visualize the learned  filters in Figure \ref{filters}. We only found two types of spatial and temporal filters from all trained \tf{} models.

\subsection{Generalization Capability}

We also did the same experiments on an additional dataset (Rayleigh number = $10^5$) to demonstrate the generalization ability of \tf{}. Figure \ref{add_results} in appendix shows the performances of TF-net, U-net and ResNet on an additional dataset. From left to right are: RMSE of different models’ predictions at varying forecasting horizon, Mean absolute divergence of models’ predictions at varying forecasting horizon, Energy Spectrum, and Turbulence kinetic energy fields of three models’ predictions. We can see that \tf{} outperforms the best two baselines, \texttt{U-net} and \texttt{ResNet} across all metrics. This demonstrate that \tf{} generalizes well to turbulent flows with a different Rayleigh number.

\section{Discussion and Future work}
We presente a novel hybrid deep learning model, \tf{}, that unifies representation learning and turbulence simulation techniques. \tf{} exploits the multi-scale behavior of turbulent flows to design trainable scale-separation operators to model different ranges of scales individually. We provide exhaustive comparisons of \tf{} and baselines and observe significant improvement in both the prediction error and desired physical quantifies, including divergence, turbulence kinetic energy and energy spectrum. We argue that different evaluation metrics are necessary  to evaluate a DL model's prediction performance for physical systems that include both accuracy and physical consistency. 
A key contribution of this work is the combination of state-of-the-art turbulent flow simulation paradigms with deep learning. Future work includes extending these techniques to very high-resolution, 3D turbulent flows and incorporating additional physical variables, such as pressure and temperature, and additional physical constraints, such as conservation of momentum, to improve the accuracy and faithfulness of deep learning models. %

\clearpage
\section*{Acknowledgement}
This research used resources of the National Energy Research Scientific Computing Center, a DOE Office of Science User Facility supported by the Office of Science of the U.S. Department of Energy under Contract No. DE-AC02- 05CH11231. The Titan Xp used for this research was donated by the NVIDIA Corporation. We thank Jared Dunnmon and Maziar Raissi for helpful discussions. We also thank Dragos Bogdan Chirila for providing the turbulent flow data.

\bibliographystyle{ACM-Reference-Format}
\bibliography{references}


\begin{thebibliography}{45}


\ifx \showCODEN    \undefined \def \showCODEN     #1{\unskip}     \fi
\ifx \showDOI      \undefined \def \showDOI       #1{#1}\fi
\ifx \showISBNx    \undefined \def \showISBNx     #1{\unskip}     \fi
\ifx \showISBNxiii \undefined \def \showISBNxiii  #1{\unskip}     \fi
\ifx \showISSN     \undefined \def \showISSN      #1{\unskip}     \fi
\ifx \showLCCN     \undefined \def \showLCCN      #1{\unskip}     \fi
\ifx \shownote     \undefined \def \shownote      #1{#1}          \fi
\ifx \showarticletitle \undefined \def \showarticletitle #1{#1}   \fi
\ifx \showURL      \undefined \def \showURL       {\relax}        \fi
\providecommand\bibfield[2]{#2}
\providecommand\bibinfo[2]{#2}
\providecommand\natexlab[1]{#1}
\providecommand\showeprint[2][]{arXiv:#2}

\bibitem[\protect\citeauthoryear{Anuj~Karpatne}{Anuj~Karpatne}{2017}]%
        {Anuj2017PGNN}
\bibfield{author}{\bibinfo{person}{Jordan Read Vipin~Kumar Anuj~Karpatne,
  William~Watkins}.} \bibinfo{year}{2017}\natexlab{}.
\newblock \showarticletitle{Physics-guided Neural Networks (PGNN): An
  Application in Lake Temperature Modeling}.
\newblock \bibinfo{journal}{\emph{arXiv Preprint arXiv:1710.11431}}
  (\bibinfo{year}{2017}).
\newblock


\bibitem[\protect\citeauthoryear{Arvind~Mohan}{Arvind~Mohan}{2019}]%
        {Arvind}
\bibfield{author}{\bibinfo{person}{Michael Chertkov Daniel~Livescu
  Arvind~Mohan, Don~Daniel}.} \bibinfo{year}{2019}\natexlab{}.
\newblock \showarticletitle{Compressed Convolutional LSTM: An Efficient Deep
  Learning Framework to Model High Fidelity 3D Turbulence}.
\newblock \bibinfo{journal}{\emph{arXiv:1903.00033}} (\bibinfo{year}{2019}).
\newblock


\bibitem[\protect\citeauthoryear{Chaoua}{Chaoua}{2017}]%
        {hybrid1}
\bibfield{author}{\bibinfo{person}{Bruno Chaoua}.}
  \bibinfo{year}{2017}\natexlab{}.
\newblock \showarticletitle{The State of the Art of Hybrid RANS/LES Modeling
  for the Simulation of Turbulent Flows}.
\newblock   \bibinfo{volume}{99} (\bibinfo{year}{2017}),
  \bibinfo{pages}{279–327}.
\newblock
\urldef\tempurl%
\url{https://doi.org/10.1007/s10494-017-9828-8}
\showDOI{\tempurl}


\bibitem[\protect\citeauthoryear{Chelsea~Finn}{Chelsea~Finn}{2016}]%
        {Chelsea}
\bibfield{author}{\bibinfo{person}{Sergey~Levine Chelsea~Finn,
  Ian~Goodfellow}.} \bibinfo{year}{2016}\natexlab{}.
\newblock \showarticletitle{Unsupervised Learning for Physical Interaction
  through Video Prediction}.
\newblock \bibinfo{journal}{\emph{arXiv:1605.07157v4}} (\bibinfo{year}{2016}).
\newblock


\bibitem[\protect\citeauthoryear{Chirila}{Chirila}{2018}]%
        {Dragos-thesis}
\bibfield{author}{\bibinfo{person}{D.~B. Chirila}.}
  \bibinfo{year}{2018}\natexlab{}.
\newblock \showarticletitle{Towards lattice boltzmann models for climate
  sciences: The GeLB programming language with application}.
\newblock  (\bibinfo{year}{2018}).
\newblock


\bibitem[\protect\citeauthoryear{Chu and Thuerey}{Chu and Thuerey}{2017}]%
        {chu2017data}
\bibfield{author}{\bibinfo{person}{Mengyu Chu} {and} \bibinfo{person}{Nils
  Thuerey}.} \bibinfo{year}{2017}\natexlab{}.
\newblock \showarticletitle{Data-driven synthesis of smoke flows with CNN-based
  feature descriptors}.
\newblock \bibinfo{journal}{\emph{ACM Transactions on Graphics (TOG)}}
  \bibinfo{volume}{36}, \bibinfo{number}{4} (\bibinfo{year}{2017}),
  \bibinfo{pages}{69}.
\newblock


\bibitem[\protect\citeauthoryear{E.~Labourasse}{E.~Labourasse}{2004}]%
        {turb3}
\bibfield{author}{\bibinfo{person}{P.~Sagaut E.~Labourasse}.}
  \bibinfo{year}{2004}\natexlab{}.
\newblock \showarticletitle{Advance in RANS-LES Coupling, a Review and an
  Insight on the NLDE Approach}.
\newblock \bibinfo{journal}{\emph{Archives of Computational Methods in
  Engineering}}  \bibinfo{volume}{11} (\bibinfo{year}{2004}),
  \bibinfo{pages}{199--256}.
\newblock


\bibitem[\protect\citeauthoryear{Emmanuel~de Bezenac}{Emmanuel~de
  Bezenac}{2018}]%
        {PDE-CDNN}
\bibfield{author}{\bibinfo{person}{Patrick~Gallinari Emmanuel~de Bezenac,
  Arthur~Pajot}.} \bibinfo{year}{2018}\natexlab{}.
\newblock \showarticletitle{Deep Learning for Physical Processes: Incorporating
  Prior Scientific Knowledge}.
\newblock \bibinfo{journal}{\emph{arXiv:1505.04597}} (\bibinfo{year}{2018}).
\newblock


\bibitem[\protect\citeauthoryear{Fang, Sondak, Protopapas, and Succi}{Fang
  et~al\mbox{.}}{2018}]%
        {fang2018deep}
\bibfield{author}{\bibinfo{person}{Rui Fang}, \bibinfo{person}{David Sondak},
  \bibinfo{person}{Pavlos Protopapas}, {and} \bibinfo{person}{Sauro Succi}.}
  \bibinfo{year}{2018}\natexlab{}.
\newblock \showarticletitle{Deep learning for turbulent channel flow}.
\newblock \bibinfo{journal}{\emph{arXiv preprint arXiv:1812.02241}}
  (\bibinfo{year}{2018}).
\newblock


\bibitem[\protect\citeauthoryear{Finn, Goodfellow, and Levine}{Finn
  et~al\mbox{.}}{2016}]%
        {finn2016unsupervised}
\bibfield{author}{\bibinfo{person}{Chelsea Finn}, \bibinfo{person}{Ian
  Goodfellow}, {and} \bibinfo{person}{Sergey Levine}.}
  \bibinfo{year}{2016}\natexlab{}.
\newblock \showarticletitle{Unsupervised learning for physical interaction
  through video prediction}. In \bibinfo{booktitle}{\emph{Advances in neural
  information processing systems}}. \bibinfo{pages}{64--72}.
\newblock


\bibitem[\protect\citeauthoryear{Izhikevich}{Izhikevich}{2007}]%
        {izhikevich2007dynamical}
\bibfield{author}{\bibinfo{person}{Eugene~M. Izhikevich}.}
  \bibinfo{year}{2007}\natexlab{}.
\newblock \bibinfo{booktitle}{\emph{Dynamical systems in neuroscience}}.
\newblock \bibinfo{publisher}{MIT press}.
\newblock


\bibitem[\protect\citeauthoryear{Jia, Willard, Karpatne, Read, Zwart,
  Steinbach, and Kumar}{Jia et~al\mbox{.}}{2019}]%
        {jia2019physics}
\bibfield{author}{\bibinfo{person}{Xiaowei Jia}, \bibinfo{person}{Jared
  Willard}, \bibinfo{person}{Anuj Karpatne}, \bibinfo{person}{Jordan Read},
  \bibinfo{person}{Jacob Zwart}, \bibinfo{person}{Michael Steinbach}, {and}
  \bibinfo{person}{Vipin Kumar}.} \bibinfo{year}{2019}\natexlab{}.
\newblock \showarticletitle{Physics guided RNNs for modeling dynamical systems:
  A case study in simulating lake temperature profiles}. In
  \bibinfo{booktitle}{\emph{Proceedings of the 2019 SIAM International
  Conference on Data Mining}}. SIAM, \bibinfo{pages}{558--566}.
\newblock


\bibitem[\protect\citeauthoryear{Jonathan~Tompson}{Jonathan~Tompson}{2017}]%
        {eulerian}
\bibfield{author}{\bibinfo{person}{Pablo Sprechmann Ken~Perlin
  Jonathan~Tompson, Kristofer~Schlachter}.} \bibinfo{year}{2017}\natexlab{}.
\newblock \showarticletitle{Accelerating eulerian fluid simulation with
  convolutional networks}. In \bibinfo{booktitle}{\emph{ICML'17 Proceedings of
  the 34th International Conference on Machine Learning}},
  Vol.~\bibinfo{volume}{70}. \bibinfo{pages}{3424--3433}.
\newblock


\bibitem[\protect\citeauthoryear{Kaiming~He}{Kaiming~He}{2015}]%
        {He2015}
\bibfield{author}{\bibinfo{person}{Shaoqing Ren Jian~Sun Kaiming~He,
  Xiangyu~Zhang}.} \bibinfo{year}{2015}\natexlab{}.
\newblock \showarticletitle{Deep Residual Learning for Image Recognition}.
\newblock \bibinfo{journal}{\emph{arXiv:1505.04597}} (\bibinfo{year}{2015}).
\newblock


\bibitem[\protect\citeauthoryear{Kim and Lee}{Kim and Lee}{2019}]%
        {kim2019deep}
\bibfield{author}{\bibinfo{person}{Junhyuk Kim} {and}
  \bibinfo{person}{Changhoon Lee}.} \bibinfo{year}{2019}\natexlab{}.
\newblock \showarticletitle{Deep unsupervised learning of turbulence for inflow
  generation at various Reynolds numbers}.
\newblock \bibinfo{journal}{\emph{arXiv:1908.10515}} (\bibinfo{year}{2019}).
\newblock


\bibitem[\protect\citeauthoryear{Li, Yu, Shahabi, and Liu}{Li
  et~al\mbox{.}}{2018}]%
        {li2018diffusion}
\bibfield{author}{\bibinfo{person}{Yaguang Li}, \bibinfo{person}{Rose Yu},
  \bibinfo{person}{Cyrus Shahabi}, {and} \bibinfo{person}{Yan Liu}.}
  \bibinfo{year}{2018}\natexlab{}.
\newblock \showarticletitle{Diffusion Convolutional Recurrent Neural Network:
  Data-Driven Traffic Forecasting}. In \bibinfo{booktitle}{\emph{International
  Conference on Learning Representations (ICLR)}}.
\newblock


\bibitem[\protect\citeauthoryear{Ling, Kurzawski, and Templeton}{Ling
  et~al\mbox{.}}{2016}]%
        {ling2016reynolds}
\bibfield{author}{\bibinfo{person}{Julia Ling}, \bibinfo{person}{Andrew
  Kurzawski}, {and} \bibinfo{person}{Jeremy Templeton}.}
  \bibinfo{year}{2016}\natexlab{}.
\newblock \showarticletitle{Reynolds averaged turbulence modeling using deep
  neural networks with embedded invariance}.
\newblock \bibinfo{journal}{\emph{Journal of Fluid Mechanics}}
  \bibinfo{volume}{807} (\bibinfo{year}{2016}), \bibinfo{pages}{155--166}.
\newblock


\bibitem[\protect\citeauthoryear{Mathieu, Couprie, and LeCun}{Mathieu
  et~al\mbox{.}}{2015}]%
        {mathieu2015deep}
\bibfield{author}{\bibinfo{person}{Michael Mathieu}, \bibinfo{person}{Camille
  Couprie}, {and} \bibinfo{person}{Yann LeCun}.}
  \bibinfo{year}{2015}\natexlab{}.
\newblock \showarticletitle{Deep multi-scale video prediction beyond mean
  square error}.
\newblock \bibinfo{journal}{\emph{arXiv preprint arXiv:1511.05440}}
  (\bibinfo{year}{2015}).
\newblock


\bibitem[\protect\citeauthoryear{Maziar~Raissi}{Maziar~Raissi}{2018}]%
        {mazier2}
\bibfield{author}{\bibinfo{person}{George Em~Karniadakis Maziar~Raissi}.}
  \bibinfo{year}{2018}\natexlab{}.
\newblock \showarticletitle{Hidden physics models: Machine learning of
  nonlinear partial differential equations}.
\newblock \bibinfo{journal}{\emph{J. Comput. Phys.}}  \bibinfo{volume}{357}
  (\bibinfo{year}{2018}), \bibinfo{pages}{125--141}.
\newblock


\bibitem[\protect\citeauthoryear{Maziar~Raissi}{Maziar~Raissi}{2019}]%
        {mazier1}
\bibfield{author}{\bibinfo{person}{George E~Karniadakis Maziar~Raissi,
  Paris~Perdikaris}.} \bibinfo{year}{2019}\natexlab{}.
\newblock \showarticletitle{Physics-informed neural networks: A deep learning
  framework for solving forward and inverse problems involving nonlinear
  partial differential equations.}
\newblock \bibinfo{journal}{\emph{J. Comput. Phys.}}  \bibinfo{volume}{378}
  (\bibinfo{year}{2019}), \bibinfo{pages}{686--707}.
\newblock


\bibitem[\protect\citeauthoryear{McDonough}{McDonough}{2007a}]%
        {turb1}
\bibfield{author}{\bibinfo{person}{J.~M. McDonough}.}
  \bibinfo{year}{2007}\natexlab{a}.
\newblock \bibinfo{booktitle}{\emph{Introductory Lectures on Turbulence}}.
\newblock \bibinfo{publisher}{Mechanical Engineering Textbook Gallery}.
\newblock
\urldef\tempurl%
\url{https://uknowledge.uky.edu/me_textbooks/2}
\showURL{%
\tempurl}


\bibitem[\protect\citeauthoryear{McDonough}{McDonough}{2007b}]%
        {mcdonough2007introductory}
\bibfield{author}{\bibinfo{person}{James~M McDonough}.}
  \bibinfo{year}{2007}\natexlab{b}.
\newblock \showarticletitle{Introductory lectures on turbulence: physics,
  mathematics and modeling}.
\newblock  (\bibinfo{year}{2007}).
\newblock


\bibitem[\protect\citeauthoryear{N.~Thuerey}{N.~Thuerey}{2019}]%
        {marie}
\bibfield{author}{\bibinfo{person}{L.~Prantl Xiangyu~Hu N.~Thuerey,
  K.~Weibenow}.} \bibinfo{year}{2019}\natexlab{}.
\newblock \showarticletitle{Deep Learning Methods for Reynolds-Averaged
  Navier-Stokes Simulations of Airfoil Flows}.
\newblock \bibinfo{journal}{\emph{arXiv:1810.08217}} (\bibinfo{year}{2019}).
\newblock


\bibitem[\protect\citeauthoryear{Olaf~Ronneberger}{Olaf~Ronneberger}{2015}]%
        {unet}
\bibfield{author}{\bibinfo{person}{Thomas~Brox Olaf~Ronneberger,
  Philipp~Fischer}.} \bibinfo{year}{2015}\natexlab{}.
\newblock \showarticletitle{U-Net: Convolutional Networks for Biomedical Image
  Segmentation}.
\newblock \bibinfo{journal}{\emph{arXiv:1512.03385}} (\bibinfo{year}{2015}).
\newblock


\bibitem[\protect\citeauthoryear{Pierre~Sagaut}{Pierre~Sagaut}{2006}]%
        {turb2}
\bibfield{author}{\bibinfo{person}{Marc~Terracol Pierre~Sagaut,
  Sebastien~Deck}.} \bibinfo{year}{2006}\natexlab{}.
\newblock \bibinfo{booktitle}{\emph{Multiscale and Multiresolution Approaches
  in Turbulence}}.
\newblock \bibinfo{publisher}{Imperial College Press}.
\newblock


\bibitem[\protect\citeauthoryear{Raissi}{Raissi}{2018}]%
        {dhpm}
\bibfield{author}{\bibinfo{person}{Maziar Raissi}.}
  \bibinfo{year}{2018}\natexlab{}.
\newblock \showarticletitle{Deep Hidden Physics Models: Deep Learning of
  Nonlinear Partial Differential Equations}.
\newblock \bibinfo{journal}{\emph{Journal of Machine Learning Research}}
  (\bibinfo{year}{2018}).
\newblock


\bibitem[\protect\citeauthoryear{Raissi, Perdikaris, and Karniadakis}{Raissi
  et~al\mbox{.}}{2017}]%
        {raissi2017physics}
\bibfield{author}{\bibinfo{person}{Maziar Raissi}, \bibinfo{person}{Paris
  Perdikaris}, {and} \bibinfo{person}{George~Em Karniadakis}.}
  \bibinfo{year}{2017}\natexlab{}.
\newblock \showarticletitle{Physics Informed Deep Learning (Part I):
  Data-driven solutions of nonlinear partial differential equations}.
\newblock \bibinfo{journal}{\emph{arXiv preprint arXiv:1711.10561}}
  (\bibinfo{year}{2017}).
\newblock


\bibitem[\protect\citeauthoryear{Reichstein, Camps-Valls, Stevens, Jung,
  Denzler, Carvalhais, and Prabhat}{Reichstein et~al\mbox{.}}{2019}]%
        {prabhat_nature_2019}
\bibfield{author}{\bibinfo{person}{Markus Reichstein}, \bibinfo{person}{Gustau
  Camps-Valls}, \bibinfo{person}{Bjorn Stevens}, \bibinfo{person}{Martin Jung},
  \bibinfo{person}{Joachim Denzler}, \bibinfo{person}{Nuno Carvalhais}, {and}
  \bibinfo{person}{Prabhat}.} \bibinfo{year}{2019}\natexlab{}.
\newblock \showarticletitle{Deep learning and process understanding for
  data-driven Earth system science}.
\newblock \bibinfo{journal}{\emph{Nature}} \bibinfo{volume}{566},
  \bibinfo{number}{7743} (\bibinfo{year}{2019}), \bibinfo{pages}{195--204}.
\newblock
\showISBNx{1476-4687}
\urldef\tempurl%
\url{https://doi.org/10.1038/s41586-019-0912-1}
\showDOI{\tempurl}


\bibitem[\protect\citeauthoryear{Rui~Wang}{Rui~Wang}{2019}]%
        {rui2020symmetry}
\bibfield{author}{\bibinfo{person}{Rose~Yu Rui~Wang, Robin~Walters}.}
  \bibinfo{year}{2019}\natexlab{}.
\newblock \showarticletitle{Incorporating Symmetry into Deep Dynamics Models
  for Improved Generalization}.
\newblock \bibinfo{journal}{\emph{ArXiv Preprint arXiv:2002.03061}}
  (\bibinfo{year}{2019}).
\newblock


\bibitem[\protect\citeauthoryear{Sagaut}{Sagaut}{2001}]%
        {LESbook}
\bibfield{author}{\bibinfo{person}{Pierre Sagaut}.}
  \bibinfo{year}{2001}\natexlab{}.
\newblock \bibinfo{booktitle}{\emph{Large Eddy Simulation for Incompressible
  Flows}}.
\newblock \bibinfo{publisher}{Springer-Verlag Berlin Heidelberg}.
\newblock
\urldef\tempurl%
\url{https://doi.org/10.1007/978-3-662-04416-2}
\showDOI{\tempurl}


\bibitem[\protect\citeauthoryear{Steffen~Wiewel}{Steffen~Wiewel}{2019}]%
        {Steffen}
\bibfield{author}{\bibinfo{person}{Nils~Thuerey Steffen~Wiewel,
  Moritz~Becher}.} \bibinfo{year}{2019}\natexlab{}.
\newblock \showarticletitle{Latent-space Physics: Towards Learning the Temporal
  Evolution of Fluid Flow}.
\newblock \bibinfo{journal}{\emph{Computer Graphics Forum}}
  \bibinfo{volume}{38} (\bibinfo{year}{2019}).
\newblock
Issue 2.


\bibitem[\protect\citeauthoryear{Steven~Brunton}{Steven~Brunton}{2019}]%
        {mlfm}
\bibfield{author}{\bibinfo{person}{Petros~Koumoutsakos Steven~Brunton,
  Bernd~Noack}.} \bibinfo{year}{2019}\natexlab{}.
\newblock \showarticletitle{Machine Learning for Fluid Mechanics}.
\newblock \bibinfo{journal}{\emph{arXiv:1905.11075}} (\bibinfo{year}{2019}).
\newblock


\bibitem[\protect\citeauthoryear{Strogatz}{Strogatz}{2018}]%
        {strogatz2018nonlinear}
\bibfield{author}{\bibinfo{person}{Steven~H. Strogatz}.}
  \bibinfo{year}{2018}\natexlab{}.
\newblock \bibinfo{booktitle}{\emph{Nonlinear dynamics and chaos: with
  applications to physics, biology, chemistry, and engineering}}.
\newblock \bibinfo{publisher}{CRC press}.
\newblock


\bibitem[\protect\citeauthoryear{Tom~Beucler}{Tom~Beucler}{2019}]%
        {constrain1}
\bibfield{author}{\bibinfo{person}{Stephan Rasp Pierre Gentine Jordan
  Ott-Pierre~Baldi Tom~Beucler, Michael~Pritchard}.}
  \bibinfo{year}{2019}\natexlab{}.
\newblock \showarticletitle{Enforcing Analytic Constraints in Neural-Networks
  Emulating Physical Systems}.
\newblock \bibinfo{journal}{\emph{arXiv:1909.00912v2}} (\bibinfo{year}{2019}).
\newblock


\bibitem[\protect\citeauthoryear{Tompson, Schlachter, Sprechmann, and
  Perlin}{Tompson et~al\mbox{.}}{2017}]%
        {tompson2017accelerating}
\bibfield{author}{\bibinfo{person}{Jonathan Tompson},
  \bibinfo{person}{Kristofer Schlachter}, \bibinfo{person}{Pablo Sprechmann},
  {and} \bibinfo{person}{Ken Perlin}.} \bibinfo{year}{2017}\natexlab{}.
\newblock \showarticletitle{Accelerating eulerian fluid simulation with
  convolutional networks}. In \bibinfo{booktitle}{\emph{Proceedings of the 34th
  International Conference on Machine Learning-Volume 70}}. JMLR. org,
  \bibinfo{pages}{3424--3433}.
\newblock


\bibitem[\protect\citeauthoryear{Villegas, Yang, Hong, Lin, and Lee}{Villegas
  et~al\mbox{.}}{2017}]%
        {villegas2017decomposing}
\bibfield{author}{\bibinfo{person}{Ruben Villegas}, \bibinfo{person}{Jimei
  Yang}, \bibinfo{person}{Seunghoon Hong}, \bibinfo{person}{Xunyu Lin}, {and}
  \bibinfo{person}{Honglak Lee}.} \bibinfo{year}{2017}\natexlab{}.
\newblock \showarticletitle{Decomposing motion and content for natural video
  sequence prediction}.
\newblock \bibinfo{journal}{\emph{arXiv:1706.08033}} (\bibinfo{year}{2017}).
\newblock


\bibitem[\protect\citeauthoryear{Wainwright and Ellis}{Wainwright and
  Ellis}{2005}]%
        {wainwright2005dynamical}
\bibfield{author}{\bibinfo{person}{John Wainwright} {and}
  \bibinfo{person}{George Francis~Rayner Ellis}.}
  \bibinfo{year}{2005}\natexlab{}.
\newblock \bibinfo{booktitle}{\emph{Dynamical systems in cosmology}}.
\newblock \bibinfo{publisher}{Cambridge University Press}.
\newblock


\bibitem[\protect\citeauthoryear{{Wu}, {Kashinath}, {Albert}, {Chirila},
  {Prabhat}, and {Xiao}}{{Wu} et~al\mbox{.}}{2019}]%
        {jinlong_2019}
\bibfield{author}{\bibinfo{person}{Jin-Long {Wu}}, \bibinfo{person}{Karthik
  {Kashinath}}, \bibinfo{person}{Adrian {Albert}}, \bibinfo{person}{Dragos
  {Chirila}}, \bibinfo{person}{{Prabhat}}, {and} \bibinfo{person}{Heng
  {Xiao}}.} \bibinfo{year}{2019}\natexlab{}.
\newblock \showarticletitle{{Enforcing Statistical Constraints in Generative
  Adversarial Networks for Modeling Chaotic Dynamical Systems}}.
\newblock \bibinfo{journal}{\emph{arXiv e-prints}} (\bibinfo{date}{May}
  \bibinfo{year}{2019}).
\newblock
\showeprint[arxiv]{physics.comp-ph/1905.06841}


\bibitem[\protect\citeauthoryear{Xie, Franz, Chu, and Thuerey}{Xie
  et~al\mbox{.}}{2018}]%
        {xie2018tempogan}
\bibfield{author}{\bibinfo{person}{You Xie}, \bibinfo{person}{Erik Franz},
  \bibinfo{person}{Mengyu Chu}, {and} \bibinfo{person}{Nils Thuerey}.}
  \bibinfo{year}{2018}\natexlab{}.
\newblock \showarticletitle{tempogan: A temporally coherent, volumetric gan for
  super-resolution fluid flow}.
\newblock \bibinfo{journal}{\emph{ACM Transactions on Graphics (TOG)}}
  \bibinfo{volume}{37}, \bibinfo{number}{4} (\bibinfo{year}{2018}),
  \bibinfo{pages}{95}.
\newblock


\bibitem[\protect\citeauthoryear{Xingjian, Chen, Wang, Yeung, Wong, and
  Woo}{Xingjian et~al\mbox{.}}{2015}]%
        {xingjian2015convolutional}
\bibfield{author}{\bibinfo{person}{SHI Xingjian}, \bibinfo{person}{Zhourong
  Chen}, \bibinfo{person}{Hao Wang}, \bibinfo{person}{Dit-Yan Yeung},
  \bibinfo{person}{Wai-Kin Wong}, {and} \bibinfo{person}{Wang-chun Woo}.}
  \bibinfo{year}{2015}\natexlab{}.
\newblock \showarticletitle{Convolutional LSTM network: A machine learning
  approach for precipitation nowcasting}. In \bibinfo{booktitle}{\emph{Advances
  in neural information processing systems}}. \bibinfo{pages}{802--810}.
\newblock


\bibitem[\protect\citeauthoryear{Xingjian~Shi}{Xingjian~Shi}{2015}]%
        {convlstm}
\bibfield{author}{\bibinfo{person}{Hao Wang Dit-Yan Yeung Wai-kin Wong
  Wang-chun~Woo Xingjian~Shi, Zhourong~Chen}.} \bibinfo{year}{2015}\natexlab{}.
\newblock \showarticletitle{Convolutional LSTM Network: A Machine Learning
  Approach for Precipitation Nowcasting}.
\newblock \bibinfo{journal}{\emph{arXiv:1506.04214}} (\bibinfo{year}{2015}).
\newblock


\bibitem[\protect\citeauthoryear{Xue, Wu, Bouman, and Freeman}{Xue
  et~al\mbox{.}}{2016}]%
        {xue2016visual}
\bibfield{author}{\bibinfo{person}{Tianfan Xue}, \bibinfo{person}{Jiajun Wu},
  \bibinfo{person}{Katherine Bouman}, {and} \bibinfo{person}{Bill Freeman}.}
  \bibinfo{year}{2016}\natexlab{}.
\newblock \showarticletitle{Visual dynamics: Probabilistic future frame
  synthesis via cross convolutional networks}. In
  \bibinfo{booktitle}{\emph{Advances in neural information processing
  systems}}. \bibinfo{pages}{91--99}.
\newblock


\bibitem[\protect\citeauthoryear{Yao, Wu, Ke, Tang, Jia, Lu, Gong, Ye, and
  Li}{Yao et~al\mbox{.}}{2018}]%
        {yao2018deep}
\bibfield{author}{\bibinfo{person}{Huaxiu Yao}, \bibinfo{person}{Fei Wu},
  \bibinfo{person}{Jintao Ke}, \bibinfo{person}{Xianfeng Tang},
  \bibinfo{person}{Yitian Jia}, \bibinfo{person}{Siyu Lu},
  \bibinfo{person}{Pinghua Gong}, \bibinfo{person}{Jieping Ye}, {and}
  \bibinfo{person}{Zhenhui Li}.} \bibinfo{year}{2018}\natexlab{}.
\newblock \showarticletitle{Deep multi-view spatial-temporal network for taxi
  demand prediction}. In \bibinfo{booktitle}{\emph{Thirty-Second AAAI
  Conference on Artificial Intelligence}}.
\newblock


\bibitem[\protect\citeauthoryear{Yi, Zhang, Wang, Li, and Zheng}{Yi
  et~al\mbox{.}}{2018}]%
        {yi2018deep}
\bibfield{author}{\bibinfo{person}{Xiuwen Yi}, \bibinfo{person}{Junbo Zhang},
  \bibinfo{person}{Zhaoyuan Wang}, \bibinfo{person}{Tianrui Li}, {and}
  \bibinfo{person}{Yu Zheng}.} \bibinfo{year}{2018}\natexlab{}.
\newblock \showarticletitle{Deep distributed fusion network for air quality
  prediction}. In \bibinfo{booktitle}{\emph{Proceedings of the 24th ACM SIGKDD
  International Conference on Knowledge Discovery \& Data Mining}}.
  \bibinfo{pages}{965--973}.
\newblock


\bibitem[\protect\citeauthoryear{Yun~Long}{Yun~Long}{2019}]%
        {HybridNet}
\bibfield{author}{\bibinfo{person}{Saibal~Mukhopadhyay Yun~Long, Xueyuan~She}.}
  \bibinfo{year}{2019}\natexlab{}.
\newblock \showarticletitle{HybridNet: Integrating Model-based and Data-driven
  Learning to Predict Evolution of Dynamical Systems}.
\newblock \bibinfo{journal}{\emph{ArXiv Preprint arXiv:1806.07439}}
  (\bibinfo{year}{2019}).
\newblock


\end{thebibliography}

\vspace{20cm}
\onecolumn
\LARGE{\textbf{Appendix}}

\begin{figure*}[htb!]
  \begin{subfigure}
    \centering  
    \includegraphics[width=0.20\textwidth]{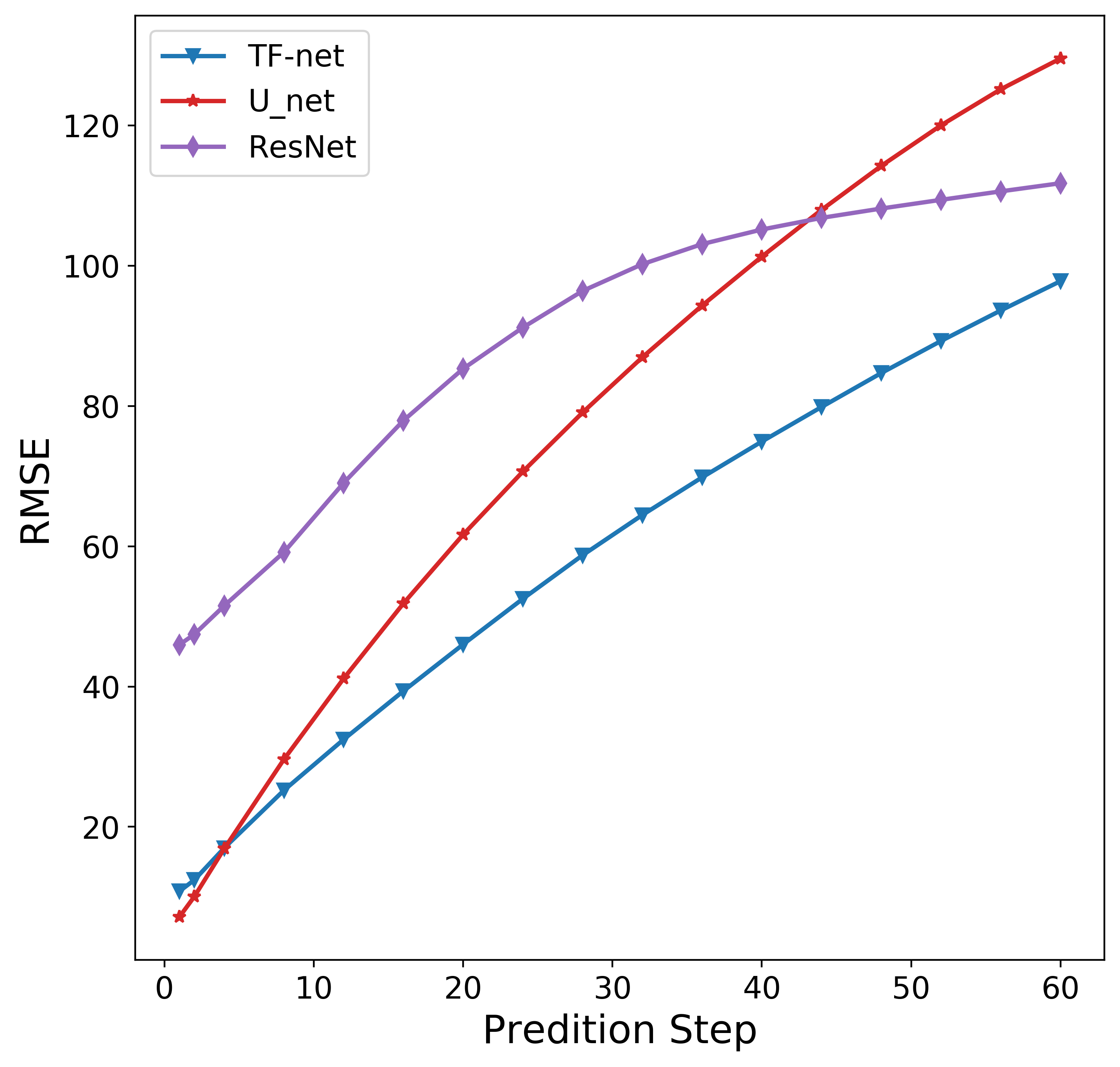}
  \end{subfigure}
  \begin{subfigure}
    \centering  
    \includegraphics[width=0.21\textwidth]{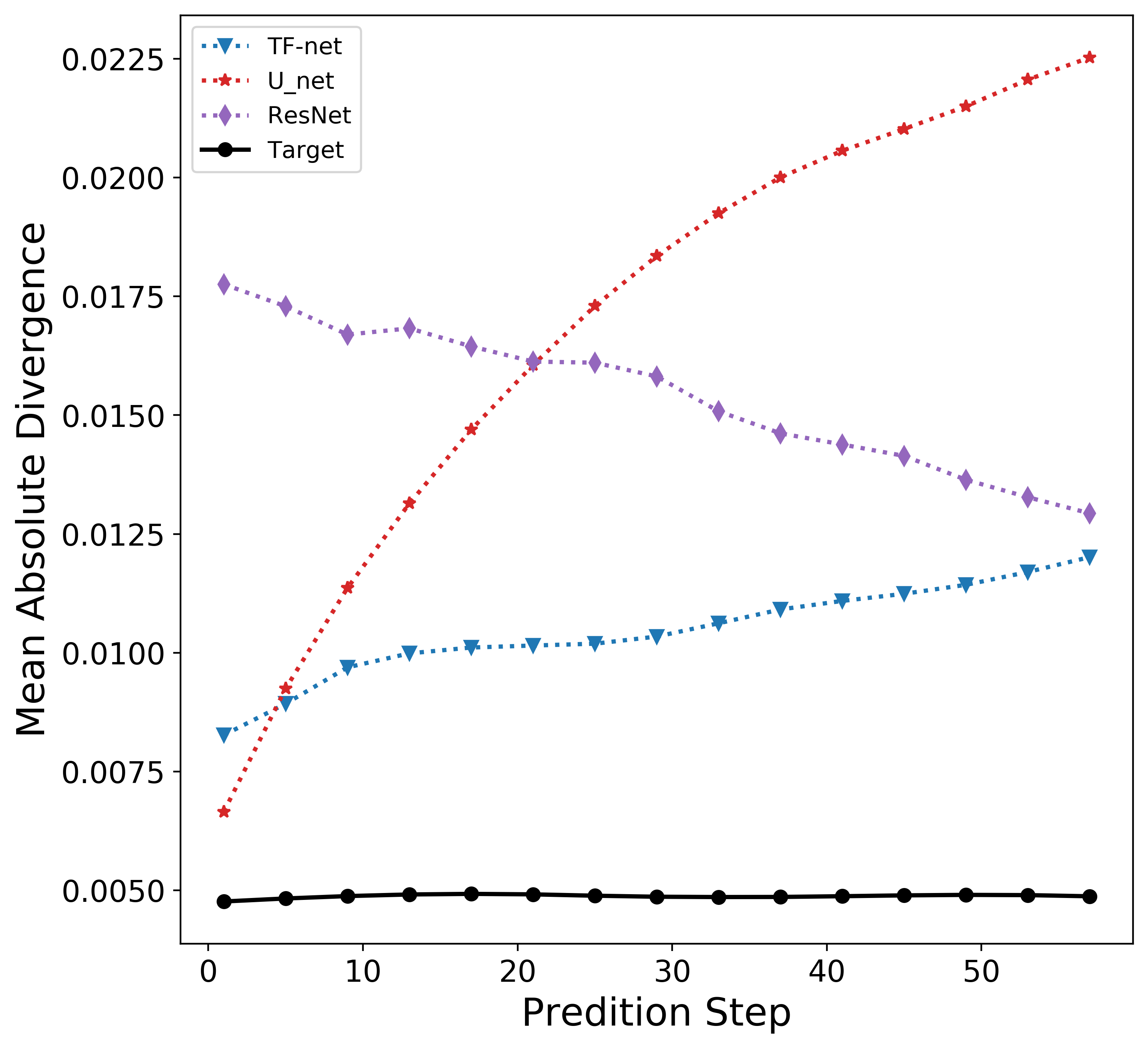}
  \end{subfigure}
  \begin{subfigure}
    \centering  
    \includegraphics[width=0.20\textwidth]{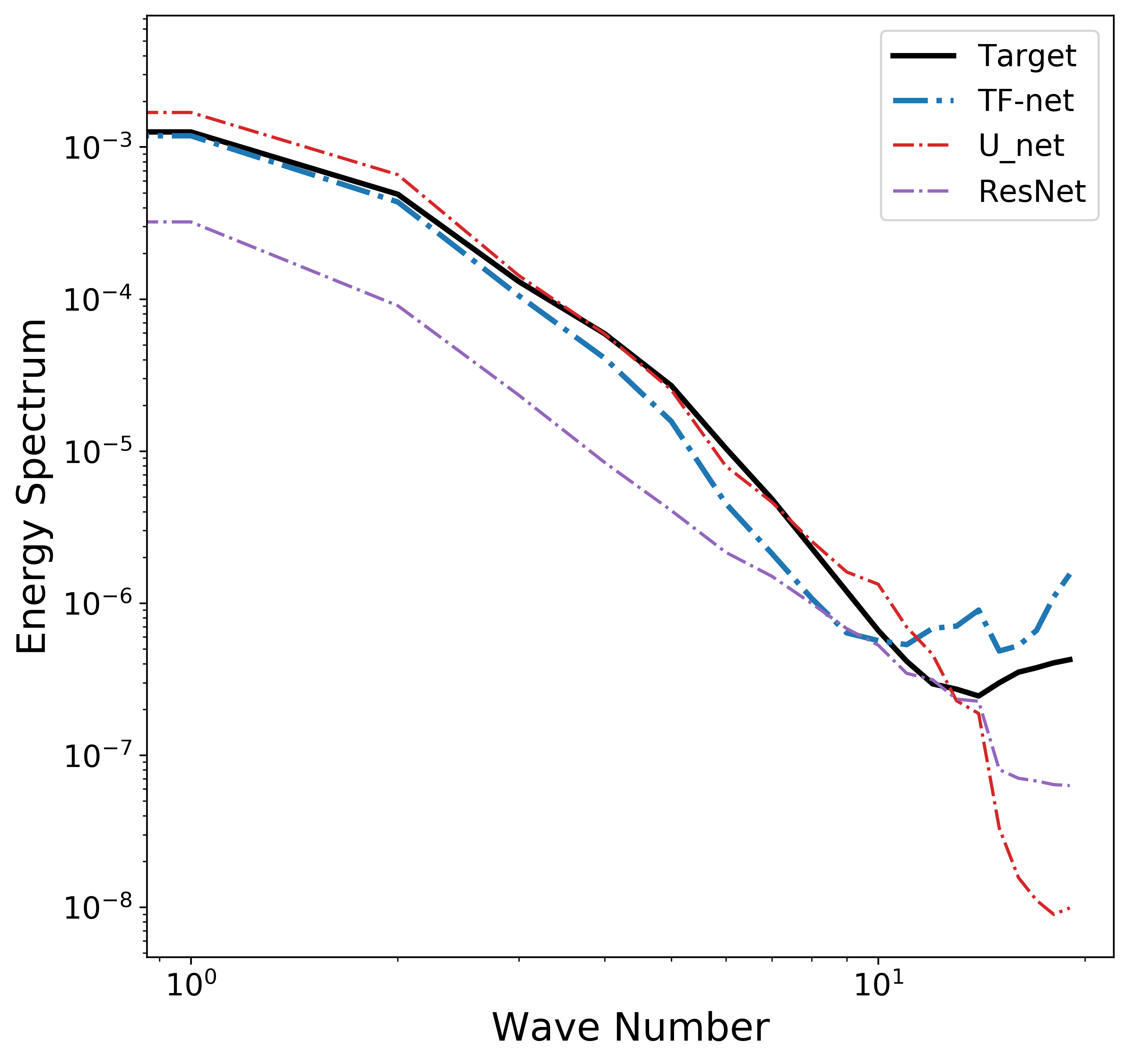}
  \end{subfigure}
  \begin{subfigure}
    \centering  
    \includegraphics[width=0.195\textwidth]{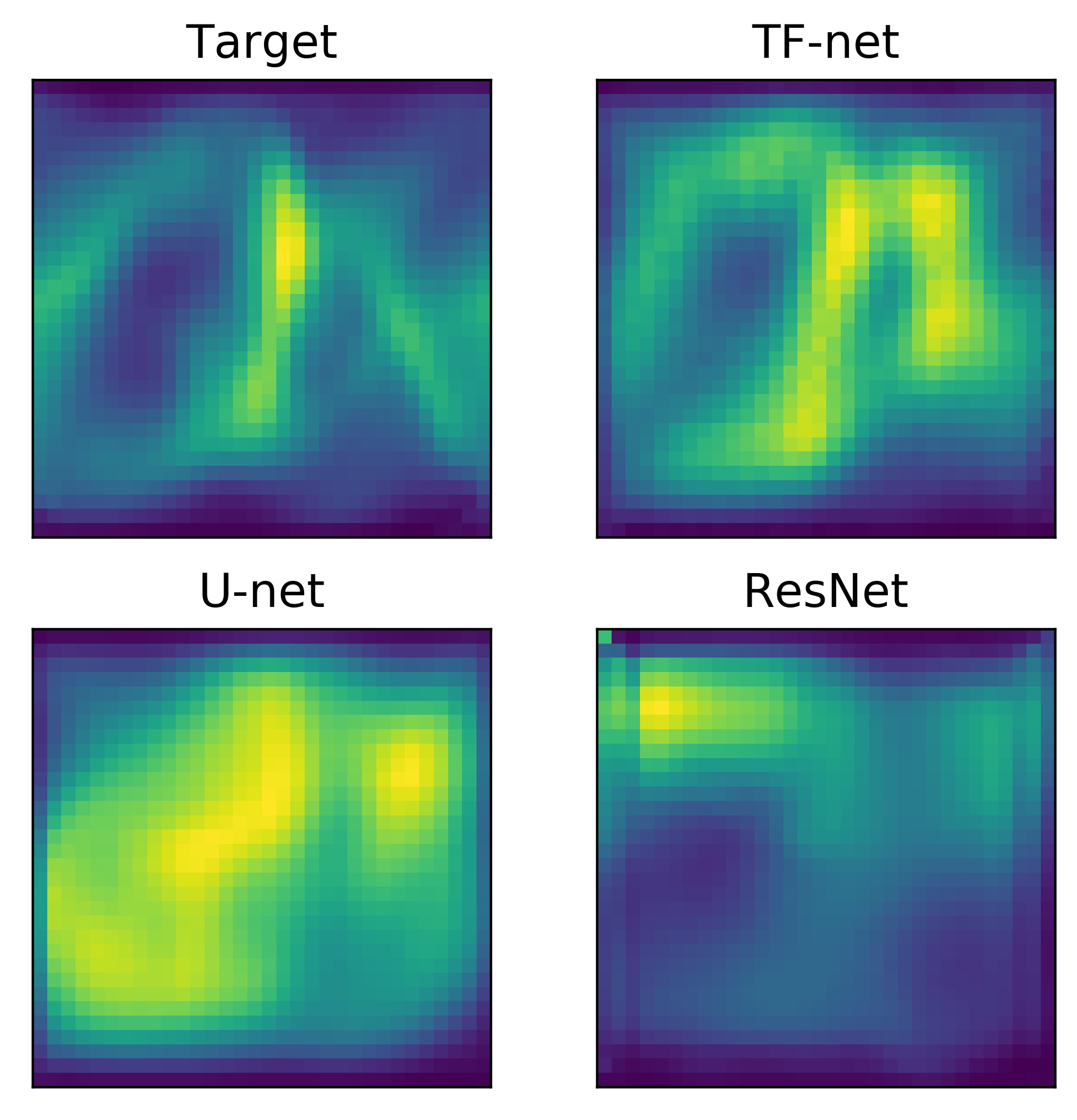}
  \end{subfigure}
\caption{The performances of TF-net, U-net and ResNet on an additional dataset(Ra = 10000). From left to right (a) Root mean square errors of different models’ predictions at varying forecasting horizon, (b) Mean absolute divergence over forecasting horizon, (c) Energy Spectrum ,and (d) Turbulence kinetic energy fields of three models’ predictions.}
\label{add_results}
\end{figure*}

\begin{table*}[hbt!]
\ra{1.3}
\normalsize
\centering
\begin{tabular}{|c|c|c|c|c|c|}
\hline
\begin{tabular}[c]{@{}c@{}}\textbf{learning rate}\end{tabular} & \begin{tabular}[c]{@{}c@{}}\textbf{Batch size}\end{tabular} & \begin{tabular}[c]{@{}c@{}}\textbf{\#Errors for backprop}\end{tabular} & \begin{tabular}[c]{@{}c@{}}\textbf{\#Input frames}\end{tabular} & \begin{tabular}[c]{@{}c@{}} \textbf{Temporal filter size} \end{tabular} & \begin{tabular}[c]{@{}c@{}} \textbf{Spatial filter size}\end{tabular} \\ \hline
1e-1 $\sim$1e-6 & 16 $\sim$128 & 1 $\sim$10 & 1 $\sim$30 & 2$\sim$10 & 3$\sim$9  \\ \hline
\end{tabular}
\caption{Hyper-parameters tuning ranges, including learning rate, batch size, the number of accumulated errors for backpropogation, the number of input frames, the moving average window size and the spatial filter size}
\label{range}
\end{table*}

\end{document}